\pdfoutput=1
\RequirePackage{ifpdf}
\ifpdf 
\documentclass[pdftex]{sigma}
\else
\documentclass{sigma}
\fi

\numberwithin{equation}{section}
\usepackage{stmaryrd}
\usepackage{delarray}
\DeclareMathOperator{\End}{End}
\DeclareMathOperator{\gen}{gen}
\DeclareMathOperator{\iso}{iso}

\begin{document}

\newcommand{\arXivNumber}{1404.2916}

\allowdisplaybreaks

\renewcommand{\thefootnote}{$\star$}

\renewcommand{\PaperNumber}{107}

\FirstPageHeading

\ShortArticleName{$\kappa$-Deformations and Extended $\kappa$-Minkowski Spacetimes}

\ArticleName{$\boldsymbol{\kappa}$-Deformations\\
and Extended $\boldsymbol{\kappa}$-Minkowski Spacetimes\footnote{This paper is a~contribution to the Special Issue on
Deformations of Space-Time and its Symmetries.
The full collection is available at \href{http://www.emis.de/journals/SIGMA/space-time.html}{http://www.emis.de/journals/SIGMA/space-time.html}}}

\Author{Andrzej BOROWIEC~$^\dag$ and Anna PACHO{\L}~$^{\ddag\S}$}

\AuthorNameForHeading{A.~Borowiec and A.~Pacho{\l}}

\Address{$^\dag$~Institute for Theoretical Physics, pl.~M.~Borna 9, 50-204 Wroc{\l}aw, Poland}
\EmailD{\href{mailto:andrzej.borowiec@ift.uni.wroc.pl}{andrzej.borowiec@ift.uni.wroc.pl}}

\Address{$^\ddag$~Science Institute, University of Iceland, Dunhaga 3, 107 Reykjavik, Iceland}
\EmailD{\href{mailto:pachol@hi.is}{pachol@hi.is}}

\Address{$^\S$~Capstone Institute for Theoretical Research, Reykjavik, Iceland}

\ArticleDates{Received April 11, 2014, in f\/inal form November 11, 2014; Published online November 22, 2014}

\Abstract{We extend our previous study of Hopf-algebraic
$\kappa$-deformations of all inhomogeneous orthogonal Lie
algebras $\iso(g)$ as written in a~tensorial and unif\/ied form.
Such deformations are determined by a~vector~$\tau$ which for Lorentzian signature can be taken time-, light- or
space-like.
We focus on some mathematical aspects related to this subject.
Firstly, we describe real forms with connection to the metric's signatures and their compatibility with the reality
condition for the corresponding $\kappa$-Minkowski (Hopf) module algebras.
Secondly, $h$-adic vs $q$-analog (polynomial) versions of deformed algebras including spe\-ciali\-za\-tion of the formal
deformation parameter~$\kappa$ to some numerical value are considered.
In the latter the general covariance is lost and one deals with an orthogonal decomposition.
The last topic treated in this paper concerns twisted extensions of $\kappa$-deformations as well as the description of
resulting noncommutative spacetime algebras in terms of solvable Lie algebras.
We found that if the type of the algebra does not depend on deformation parameters then specialization is possible.}

\Keywords{quantum deformations; quantum groups; quantum spaces; reality condition for Hopf module algebras; $q$-analog
and specialization versions; $\kappa$-Minkowski spacetime; extended $\kappa$-deformations; twist-deformations;
classif\/ication of solvable Lie algebras}

\Classification{81T75; 58B22; 16T05; 17B37; 81R60}

\renewcommand{\thefootnote}{\arabic{footnote}}
\setcounter{footnote}{0}

\section{Introduction}
Recently we have proposed a~unif\/ied description for Drinfel'd type quantization of inhomogeneous orthogonal algebras
$\iso(g)$~\cite{BP_unif} corresponding to the noncommutative spacetime of the form:
$[x^{\mu},x^{\nu}]=\frac{i}{\kappa}(\tau^{\mu}x^{\nu}-\tau^{\nu}x^{\mu})$~\cite{koma95,LLM} (we
shall call it $\kappa (\tau)$-Minkowski spacetime).
In this paper we want to discuss related problems with this generalized $\kappa(\tau)$-deformation.
Historically, the $\kappa$-Minkowski spacetime~\cite{MR, Z} was the f\/irst example of a~noncommutative spacetime of the
Lie-algebraic type, where the time coordinate does not commute with the space coordinates.
It is the time-like ($\tau^{0}\neq0$) version of above, more general noncommutative spacetime.
The $\kappa$-Minkowski spacetime was inspired by the introduction of the deformed Poincar\'{e} Hopf algebra in
1991~\cite{LNR, Luk1} with the deformation parameter 'kappa' of mass dimension (which is usually interpreted as Planck
mass or quantum gravity scale).
Lie algebraic noncommutativity, including $\kappa$-Minkowski, have been further investigated by many authors in the
wide range of applications, among others in deformation of special relativity framework~\cite{dsr1,dsr2}, noncommutative
f\/ield theo\-ries~\cite{Wess2,Wess3, ncqft1,ncqft3,ncqft4,Wess1,ncqft2,ncqft6,ncqft7,ncqft8,ncqft5}, deformed
statistics~\cite{kstatistics1,kstatistics3,kstatistics4,kstatistics2}, Planck scale
physics~\cite{planck_scale1,planck_scale2} and quantum gravity
phenomenology~\cite{qgphenom2,qgphenom3,qgphenom1,qgphenom4}.
Moreover due to the fact that in the context of $(2+1)$-dimensional quantum gravity noncommutative spacetime geometry and
deformations of Poincar\'{e} symmetry arise naturally, recently the $\kappa$-deformation in $(2+1)$ dimensions has
attracted quite some attention as well, see
e.g.~\cite{Ballesteros3dim1,Ballesteros3dim2,Ballesteros3dim3, Matschull,Schroers-Meusb}.
In~\cite{BP_unif} the deformation of the symmetry algebra (i.e.\ generalized for any $\tau$ the $\kappa$-deformed
inhomogeneous orthogonal Hopf algebra $U_{\kappa,\tau}(\iso(g))$) was determined by a~metric
tensor~$g$ of any dimension and arbitrary signature.
Such formulation allows for many applications within the deformed general relativity and Planck scale physics, including $(2+1)$-dimensional
case allo\-wing for relation with quantum gravity models.
The deformation of inhomogeneous orthogonal algebras corresponding to $\kappa (\tau)$-Minkowski spacetime explicitly
depends on the choice of an additional vector f\/ield $\tau$ which at the same time parameterizes classical $r$-matrices
and distinguishes between nonequivalent deformations\footnote{For the Lorentzian signature we can distinguish three
(nonequivalent) Hopf-algebraic deformations: time-like, space-like (a.k.a.\ tachyonic) and light-like (a.k.a.\
light-cone) quantizations of the Poincar\'{e} algebra, represented by dif\/ferent choices of vector $\tau$, i.e.\
$(1,0,0,0)$, $(1,0,0,1)$ and $(0,0,0,1)$ respectively, provided that the metric is in diagonal form:
$g_{\mu\nu}=(-,+,+,+)$.}.

In this paper we want to focus on mathematical issues connected with this generalized $\kappa (\tau)$- deformation and
discuss related problems.
The framework of Drinfel'd type of deformations~\cite{Drinfeld1,Drinfeld3} requires the so-called $h$-adic
topology~\cite{ChP,Klimyk}, i.e.\ dealing with formal power series therefore one usually uses in fact topological
completion $U(\iso(g))[[\frac{1}{\kappa}]]$.
The version of $\kappa$-Poincar\'{e} algebra used by many authors~\cite{LNR,Luk1,Luk2, MR} in a~traditional approach is
(implicitly) of such `$h$-adic' type.
In this approach the deformation parameter $\kappa$ cannot take a~numerical value and must stay formal which makes it
dif\/f\/icult to undergo some physical interpretation, e.g., as Planck mass or quantum gravity scale.
Nevertheless there exists a~method which in some cases allows deformation parameter to take a~constant (numerical)
value.
In this sense the $\kappa$-Poincar\'{e} quantum group with $h$-adic topology as described
before~\cite{koma95,LLM,LNR,Luk1,Luk2,MR,Z}, is not the only possible version.
In this paper we will reformulate this traditional Hopf algebra to hide the inf\/inite series on the abstract level.
This is known to be always possible in the framework of Drinfel'd--Jimbo standard quantization of semi-simple Lie
algebras.
As a~f\/irst step, one distinguishes certain sub-Hopf algebra which can be treated as a~Hopf algebra over polynomial ring
$\mathbb{C}[q,q^{-1}]$ (the so called `$q$-analog' version), in this way, getting rid of\/f $h$-adic topology.
Next we create new Hopf algebra by setting up (specializing) the formal parameter~$q$ to some numerical (complex in
general) value.
As a~result, one obtains a~one-parameter family of new Hopf algebras labeled by the numerical parameter which (when
real) could be interpreted as some physical quantity.
Usually, the value of the parameter matters and may lead to non-isomorphic Hopf algebras (e.g., for roots of unity) or
inf\/luence representation theory.
In our case a~real value for this parameter is dictated by real forms of (complex) Hopf algebras under consideration
which can be extended to module algebras as well.
These issues are studied with more care.

In the following we shall introduce the '$q$-analog' version of $\kappa (\tau)$-deformation of non-semi\-simple
inhomogenous orthogonal algebras $\iso(g)$, for all values of the vector~$\tau$ with a~f\/ixed deformation parameter,
i.e.~$\kappa\in\mathbb{C}$.
It appears that in this case one can re-scale all the formulas to get rid of\/f the deformation parameter altogether.
Firstly we recall the case of the orthogonal $D=1+(D-1)$ decomposition for time-like~$\tau$ and introduce
its $q$-analog version (analogous arguments will hold for space-like case as well).
Later we consider the null-plane deformation and $D=2+(D-2)$ decomposition.
One should underline that the $q$-analog version has been common for the standard deformations (i.e.~for
which $r$-matrix satisf\/ies modif\/ied Yang--Baxter equation (MYBE), as in the time-like case).
On the contrary it is rather unexpected for non-standard deformations (i.e.~for which $r$-matrix satisf\/ies classical YBE
and when the twist exists, like in the light-like case).
As a~by-product we show that specialization procedure can be also applied to the extended Jordanian twist deformations.
Also the underlying covariant quantum space, together with its real form, is introduced in the $q$-analog as well as
specialized version.
Such $\kappa(\tau)$-Minkowski spacetime with f\/ixed value of parameter~$\kappa$ is an universal envelope of solvable Lie
algebra without $h$-adic topology.
This version has been already considered by some authors, e.g.\ in the context of spectral triples~\cite{Andrea,SitDur,Sitarz3,Sitarz2,Sitarz1}
and group f\/ield theory~\cite{Oriti}.

As a~f\/inal issue related with $\kappa(\tau)$-deformation, we shall deal in this paper, is the possibility of extending
the $\kappa$-deformations via twisting.
Quantum deformations for Lorentz and Poincar\'{e} symmetries have been classif\/ied in terms of
classical $r$-matrices~\cite{Zakrz-Comm}.
For example the original $\kappa$-deformation of Poincar\'{e} algebra corresponds to $r=M_{0i}\wedge P^i$.
It is a~particular case of more general family $r_\tau=\tau^\alpha M_{\alpha\mu}\wedge P^\mu$ found for any non-zero
vector~$\tau$ by Zakrzewski~\cite{Zakrz-Comm}.
It has been shown in the same paper that $r_\tau$ admits extensions $r_\tau+ \xi r^{\prime}$ with an additional
parameter~$\xi$, where $r^{\prime}$ corresponds to some specif\/ic triangular deformations.
The passage from a~classical $r$-matrix to the corresponding deformation is, in general, non trivial task even in
a~triangular case: as an intermediate step one needs to construct a~twisting element before quantizing.
Twisting two-tensors corresponding to Zakrzewski's scheme have been already found in~\cite{Tol0704.0081a,Tol0704.0081b}
(see also \cite{BLT4,BLT3,BLT2,BLT1, LL-0406155,Lyakhovsky}).
Here we use such twists (corresponding to the extended classical $r$-matrices) in order to deform
$\kappa(\tau)$-Minkowski spacetime algebra and describe resulting algebras in terms of solvable Lie algebras.
To this aim we use classif\/ication scheme of low-dimensional solvable Lie algebras proposed in~\cite{Graaf}.
As a~f\/inal task in this paper we calculate the deformed coproducts for some selected twistings of
the $\kappa$-deformations.
It turns out that the possibility of their specialization is related with the type of twisted $\kappa$-Minkowski
spacetime algebras.

\section[$\kappa(\tau)$-deformations for (inhomogeneous) orthogonal Lie algebras]{$\boldsymbol{\kappa(\tau)}$-deformations
for (inhomogeneous) orthogonal\\ Lie algebras}

Here we recall the form of $\kappa$-deformations recently written in an unif\/ied way~\cite{BP_unif}.
Let~$V$ be a~$D$-dimensional (real) vector space equipped with a~metric tensor~$g$ of arbitrary signature $(p, q)$, $p+q=D$.
For an arbitrary basis $\{e_{\mu}\}_{\mu=0}^{D-1}$ one can introduce its components $g_{\mu
\nu}=g(e_{\mu},e_{\nu})$.
Making use of the dual basis $\{e^{\mu}\}_{\mu=0}^{D-1}$ in the dual vector space $V^\#$ one can write
$g=g_{\alpha\beta}e^\alpha\otimes e^\beta$.
It is well-known that Lie algebra of inhomogeneous orthogonal group ${\rm ISO}(g)$ consists of ${\frac{1}{2}}D(D+1)$
generators $(M_{\mu\nu}, P_\alpha)$ adapted to a~choice of the basis and satisfying the standard commutation relations
\begin{gather}
[M_{\mu \nu},M_{\rho \lambda}]=i(g_{\mu \lambda}M_{\nu \rho}-g_{\nu \lambda}M_{\mu \rho}+g_{\nu \rho}M_{\mu
\lambda}-g_{\mu \rho}M_{\nu \lambda}),
\label{MM}
\\
[M_{\mu \nu},P_{\rho}]=i(g_{\nu \rho}P_{\mu}-g_{\mu \rho}P_{\nu}),
\qquad
[P_{\mu},P_{\lambda}]=0.
\label{PP}
\end{gather}
The relation with the basis $\{e_{\mu}\}_{\mu=0}^{D-1}$ of~$V$ is throughout the complexify vector representation\footnote{Throughout this paper we shall use the standard covariant Einstein's convention under which the repeated
covariant and contravariant indices indicate summation, as well as the possibility of lowering and rising indices by the
metric $g_{\alpha\beta}$ and its inverse $g^{\alpha\beta}$.}
\begin{gather}
\label{rep}
M_{\mu\nu}\mapsto -i (g_{\mu\alpha}e_\nu-g_{\nu\alpha}e_\mu)\otimes e^\alpha \in \End V\otimes \mathbb{C}
\end{gather}
acting in the complexif\/ied vector space $V\otimes \mathbb{C}$.
In fact, the generators $(M_{\mu\nu}, P_\alpha)$ belong to the complexif\/ied Lie algebra $\iso(g)$.
However, for the purpose of this paper we shall treat the relations~\eqref{MM}, \eqref{PP}
as generating ones for
a~complex universal enveloping algebra $U(\iso(g))$ of $\iso(g)$ understood as a~free (complex)
unital associative algebra generated by the symbols $(M_{\mu\nu}, P_\alpha)$ and factorized further by a~two-sided ideal
generated by the relations~\eqref{MM},~\eqref{PP}.
It is also customary to consider real algebras as complex ones equipped additionally with the structure of involutive
anti-linear anti-automorphism, the so-called $*$-conjugation: $X\mapsto X^*$, i.e.\ having the same formal properties as
a~Hermitean conjugation.
In our case the choice of real structure is completely determined by the requirement that the
generators~\eqref{MM},~\eqref{PP}
are self-conjugated (formally Hermitean or self-adjoint), i.e.
\begin{gather}
\label{RC1}
X=X^*
\qquad
\text{for}
\quad
X\in (M_{\mu\nu}, P_\alpha)
\end{gather}
since the relations~\eqref{MM},~\eqref{PP}
are invariant with respect to such conjugation\footnote{If one wishes to have
a~real instead pure imaginary structure constants in~\eqref{MM},~\eqref{PP}
then it is necessary to re-scale the
generators $X\mapsto \tilde X=-i X$.
In such a~case $\tilde X^*=-\tilde X$ are real.}.
It is well-known that real structures def\/ined in this way are in one-to-one correspondence with the metric signatures~$(p, q)$, however complexif\/ied algebra is signature independent.

It has been found in~\cite{Zakrz-Comm} that for any (non-zero) vector $\tau=\tau^{\mu}e_{\mu}\in V$ (and any metric
tensor~$g_{\mu\nu}$ as above) one can introduce the corresponding classical $r$-matrix
\begin{gather}
r_{(\tau,g)}={\tau}^{\alpha}M_{\alpha \mu}\wedge P^{\mu}\equiv {\tau}^{\alpha}g^{\beta\sigma}M_{\alpha \beta}\wedge
P_{\sigma} \equiv {\frac{1}{2}}\tau\llcorner \Omega_g \in \wedge^{2}\iso(g),
\end{gather}
where $\Omega_g=M_{\mu \nu}\wedge P^{\mu}\wedge P^{\nu}$ is known to be the only invariant element in
$\wedge^{3}\iso(g)$ and $\tau \llcorner$ is used for contraction with the vector $\tau$.
The Schouten bracket reads
\begin{gather}
[[r_{(\tau,g)},r_{(\tau,g)}]]=- \tau_g^2\Omega_g,
\label{z2}
\end{gather}
where $\tau_g^2\equiv \tau^\mu\tau_\mu\equiv g_{\mu\nu}\tau^\mu\tau^\nu$ denotes the scalar square of~$\tau$ with
respect to the metric~$g$.
One should notice that the case $\tau_g^2=0$, which is only possible for non-Euclidean signature, provides a~solution of
the classical (non-modif\/ied) Yang--Baxter equation.
Further on we shall simplify the notation and drop of\/f the sub-index referring to the metric~$g$.

We are now in position to introduce the corresponding quantized Hopf algebra structure.
According to~\cite{Zakrz-Comm} non-equivalent quantizations are classif\/ied by the conjugation classes of the stability
subgroups $G_\tau$ of the vector~$\tau$\footnote{In fact, Zakrzewski has provided a~classif\/ication of the
classical $r$-matrices in which the modif\/ied Yang--Baxter case is not completed.
Therefore $\kappa$-deformation can be further quantized.
We shall return to this point in the last two sections.}.
For a~non-Euclidean metric it provides three non-isomorphic cases, which for the Lorentzian signature are the very
well-known ones: vector $\tau$ can be time-, space- or light-like.
In the Euclidean case there is only one $\kappa$-deformation\footnote{Euclidean case has been also studied before
in~\cite{Wess1,Meljanac0702215}.}.
In the complex (signature independent) case one distinguishes two subcases instead: $\tau^2=0$ and $\tau^2\neq0$.

In the Drinfel'd quantization scheme one requires the so-called $h$-adic topology: extension of $U(\iso(g))$
by formal power series $U(\iso(g)) [[\frac{1}{\kappa}]]$ in order to arrange the deformation.
It enables, e.g., existence of invertible twist, etc.\
(see, e.g.,~\cite{ChP, Drinfeld1,Drinfeld3} for more details).
Hereafter for shortening the notation one introduces the following objects
\begin{gather}
\Pi_{\tau}=\frac{1}{\kappa}P_{\tau}+\sqrt{1+\frac{\tau^{2}}{\kappa^{2}}C},
\qquad
\Pi_{\tau}^{-1}=\frac{\sqrt{1+\frac{\tau^{2}}{\kappa^{2}}C}-\frac{1}{\kappa}P_{\tau}}{1+\frac{1}{\kappa^{2}}\left(\tau^{2}C-P_{\tau}^{2}\right)},
\\
\tau^{2}C_{\tau}=\kappa^{2}\left(\Pi_{\tau}+\Pi_{\tau}^{-1}-2+\frac{1}{\kappa^{2}}\left(\tau^{2}C-P_{\tau}^{2}\right) \Pi_{\tau}^{-1}\right)
=2\kappa^{2}\left(\sqrt{1+\frac{\tau^{2}}{\kappa^{2}}C} -1\right)
\end{gather}
as formal power series in ${\frac{1}{\kappa}}$, where $P_{\tau}=\tau^{\mu}P_{\mu}$.
Moreover $C\equiv P^{\alpha}P_{\alpha}=g^{\alpha \beta}P_{\alpha}P_{\beta}$ denotes the well-known quadratic Casimir
element (a.k.a.\
Casimir of mass in $D=4$ Lorentzian case).
The element $C_{\tau}$ is also central and plays a~role of deformed Casimir describing deformed dispersion relations
(see, e.g.,~\cite{planck_scale1,planck_scale2} and references therein).
For the case $\tau^{2}=0$ one should take $C_{\tau}=C$.
With this notation $\kappa(\tau)$-deformed (inhomogeneous) orthogonal Lie algebra, besides the standard orthogonal Lie
algebra structure~\eqref{MM},~\eqref{PP},
has deformed coalgebraic sector~\cite{BP_unif} (cf.\ realization dependent form
in~\cite{Wess1,1110.0944,Meljanac0702215})
\begin{gather}
\Delta_{\tau}(P_{\mu})=P_{\mu}\otimes \Pi_{\tau}+1\otimes
P_{\mu}-\frac{\tau_{\mu}}{\kappa}P^{\alpha}\Pi_{\tau}^{-1}\otimes
P_{\alpha}-\frac{\tau_{\mu}}{2\kappa^{2}}C_{\tau}\Pi_{\tau}^{-1}\otimes P_{\tau},
\label{copP}
\\
\Delta_{\tau}(M_{\mu \nu})=M_{\mu \nu}\otimes 1+1\otimes M_{\mu
\nu}+\frac{1}{\kappa}P^{\alpha}\Pi_{\tau}^{-1}\otimes (\tau_{\nu}M_{\alpha \mu}-\tau_{\mu}M_{\alpha \nu})
\nonumber
\\
\phantom{\Delta_{\tau}(M_{\mu \nu})=}
 - \frac{1}{2\kappa^{2}}C_{\tau}\Pi_{\tau}^{-1}\otimes (\tau_{\mu}M_{\tau \nu}-\tau_{\nu}M_{\tau \mu}),
\label{copM}
\end{gather}
where $M_{\tau \lambda}=\tau^{\alpha}M_{\alpha \lambda}$ and $\tau_{\mu}=g_{\alpha \mu}\tau^{\mu}$ denote covariant
components of $\tau^{\mu}$ with respect to the metric $g_{\mu \nu}$.
In order to complete the Hopf algebra structure one def\/ines counits: $\epsilon (1)=1$, $\epsilon (P_{\mu})=-P_{\mu}$,
$\epsilon (M_{\mu \nu})=-M_{\mu \nu}$ and antipodes
\begin{gather}
S_{\tau} (P_{\mu} )
=-\left(P_{\mu}+\frac{\tau_{\mu}}{\kappa}\left(C+\frac{1}{2\kappa}P_{\tau}C_{\tau}\right)\right) \Pi_{\tau}^{-1},
\qquad
S_{\kappa}(\Pi_{\tau})=\Pi_{\tau}^{-1},
\label{SP}
\\
S_{\tau} (M_{\mu \nu} )=-M_{\mu \nu}+\frac{1}{\kappa} P^{\alpha} (\tau_{\nu}M_{\alpha
\mu}-\tau_{\mu}M_{\alpha \nu} ) +\frac{1}{2\kappa^{2}}C_{\tau} (\tau_{\nu}M_{\tau \mu}-\tau_{\mu}M_{\tau
\nu} ).
\label{SM}
\end{gather}
So def\/ined Hopf algebra structure will be denoted as $U_{\kappa,\tau}(\iso(g))$, i.e., in
particular, $U_{\kappa,\tau}(\iso(g))$ $\cong U(\iso(g))[[\frac{1}{\kappa}]]$
as an algebra.
It preserves the reality condition (\ref{RC1}) induced by the metric signature in the following well-known Hopf algebra
reality condition form (see e.g.~\cite{Majid-book})
\begin{gather}
\Delta_{\tau}(X^{\ast})=\Delta_{\tau}(X)^{\ast \otimes \ast}
\qquad
\text{and}
\qquad
S_{\tau}(S_{\tau}(X^{\ast})^{\ast})=X
\label{RC2}
\end{gather}
for arbitrary $X\in U_{\kappa,\tau}(\iso(g))$ provided that the vector $\tau^{\mu}$ and the
formal parameter ${\frac{1}{\kappa}}$ are real.
Above conditions are enough to be checked on self-adjoint generators~\eqref{MM},~\eqref{PP}.
In particular, the identity: $S_{\tau}(S_{\tau}(X))=\Pi_{\tau}^{D-1}X\Pi_{\tau}^{1-D}$ found in~\cite{1110.0944} can be
very helpful.

It is important to notice that simultaneous re-scaling of~$\tau$ and~$\kappa$ by the same factor does not change
formulas~\eqref{copP}--\eqref{SM} involving these symbols, so it can be treated as an isomorphism of the corresponding
Hopf algebras, i.e.~$U_{\kappa,\tau}(\iso(g))\cong U_{\lambda\kappa,\lambda\tau}(\iso(g))$.
As a~practical application one f\/inds that the vector $\tau$ can be normalized to the values $\tau^2=\pm 1, 0$.

This unif\/ied description has the advantage of general covariance manifested via tensorial character of all def\/ining
formulas~\eqref{MM}--\eqref{SM}.
Consider a~change of basis in the space~$V$: $e_{\mu}\mapsto \tilde{e}_{\mu}=A_{\mu}^{\alpha}e_{\alpha}$~by
a~non-degenerate matrix $A_{\alpha}^{\beta}\in {\rm GL}(D,\mathbb{R})$.
Thus one can introduce the new generators
\begin{gather}
\tilde{P}_{\alpha}=A_{\alpha}^{\mu}P_{\mu},
\qquad
\tilde{M}_{\alpha \beta}=A_{\alpha}^{\mu}A_{\beta}^{\nu}M_{\mu \nu}
\label{cov}
\end{gather}
together with $\tilde{g}_{\alpha \beta}=A_{\alpha}^{\mu}A_{\beta}^{\nu}g_{\mu \nu}$,
$\tilde{\tau}_{\alpha}=A_{\alpha}^{\mu}\tau_{\mu}$ (but $\tilde{\tau}^{\alpha}=(A^{-1})_{\mu}^{\alpha}\tau^{\mu}$,
$\tilde{C}=C$ and therefore \mbox{$P_{\tilde{\tau}}=P_{\tau}$}).
Then all formulas of this section remain valid if we replace all objects without tilde sign by the corresponding ones
with the tilde.
Moreover the real structure~\eqref{RC1} is preserved.
It means that $U_{\kappa,\tau}(\iso(g))\cong
U_{\kappa,\tilde{\tau}}(\iso(\tilde{g}))$ as real Hopf algebras.
In particular, if $A_{\alpha}^{\beta}\in O(g)$ then \mbox{$g_{\alpha \beta}=\tilde{g}_{\alpha \beta}$} (internal automorphism).
This fact is important for possible physical applications and interpretations (see last section in~\cite{BP_unif}).
It is to be observed that the transformation~\eqref{cov} does not change the metric signature.

\section[Specialization of the $\kappa(\tau)$-inhomogeneous orthogonal Hopf algebras]{Specialization
of the $\boldsymbol{\kappa(\tau)}$-inhomogeneous orthogonal\\ Hopf algebras}

It is known that quantized enveloping Lie algebras have many (non-isomorphic) incarnations, with the deformation
parameter being both a~formal variable (algebraic generator) or a~numerical factor.
The passage from the topological $h$-adic (Drinfel'd) version, we have used till now, to the $q$-analog (Drinfel'd--Jimbo)
form and f\/inal specialization of the formal deformation parameter to some numerical value is well understood and
described for standard deformations of semi-simple Lie algebras (see, e.g.,~\cite[Chapter~9]{ChP}
and~\cite[Chapters~3,~7]{Klimyk}).
More deeper mathematical study of this problem can be found in~\cite{Bonneau2, Bonneau1}.
The specialization problem for time-like $\kappa$-Poincar\'{e} case has been also treated in a~bicrossproduct basis in
the context of possible $C^*$-algebra reformulation~\cite{Andrea,Sitarz3,Sitarz2,Stachowiak}.

In the present section we recall our earlier result concerning specialization problem
for a~time-like version of $\kappa$-type deformation~\cite{BP2}.
Then we extend the analysis to the non-standard (light-cone) case as well as to twisted $\kappa$-deformations.
Surprisingly, to our best knowledge, the specialization problem for non-standard (i.e.~invoked by a~two-cocycle twist)
deformation has not been studied in the literature yet.
Our results show that (extended) Jordanian twist, in contrast to the Abelian one, enables (after suitable change of
variables) to solve specialization problem.

\subsection[The $D=1+(D-1)$ orthogonal decomposition: $\tau^2\neq0$]{The $\boldsymbol{D=1+(D-1)}$ orthogonal decomposition: $\boldsymbol{\tau^2\neq0}$}

In order to introduce the $q$-analog version of $\kappa (\tau)$-quantized inhomogeneous orthogonal Hopf algebra from
the previous section we start with reminding the orthogonal $D=1+(D-1)$ decomposition which relays on suitable
change\footnote{It can be done by certain choice of the basis $\{e_{\mu}\}_{\mu=0}^{D-1}$ in~$V$, where $e_{0}=\tau$
and $\{e_{i}\}_{i=1}^{D-1}$ are orthogonal to $\tau$: $g_{00}=\tau^{2}$,
$g_{0i}=g(e_{0},e_{i})=0$.
This can be reached by an analog of the so-called Gram--Schmidt orthogonalization procedure which provides the
orthogonal decomposition $(V,g_{\mu \nu})\cong (\mathbb{R},g_{00})\times (V^{D-1},g_{ij})$.
One can note that the $(D-1)$-dimensional
metric $g_{ij}$ does not need to be in the diagonal form.} of generators in
$U(\iso(g))$, provided that $\tau^{2}\neq0$.

In the corresponding Lie algebra basis $\{P_{\tau},P_{i},M_{\tau i},M_{ij}\}$ the algebraic relations read as
\begin{gather}
[M_{ij},M_{kl}]=i(g_{il}M_{jk}-g_{jl}M_{ik}+g_{jk}M_{il}-g_{ik}M_{jl}),\label{orthDec}
\\
[M_{\tau j},M_{kl}]=i(g_{jk}M_{\tau l}-g_{jl}M_{\tau k}),
\qquad
[M_{\tau j},M_{\tau l}]=0,
\\
[P_{\tau},P_{k}]=[P_{i},P_{j}]=[P_{\tau},P_{\tau}]=0,
\\
[M_{ij},P_{k}]=i(g_{jk}P_{i}-g_{ik}P_{j}),
\qquad
[M_{ij},P_{\tau}]=0,
\\
[M_{\tau j},P_{k}]=ig_{jk}P_{\tau},
\qquad
[M_{\tau j},P_{\tau}]=-i\tau^{2}P_{j},
\end{gather}
while the coproducts take the form (notice that now $\tau^\mu=(1,0,\dots,0)$)
\begin{gather}
\Delta_{\tau} (P_{\tau} )=P_{\tau}\otimes
\left(\frac{1}{\kappa}P_{\tau}+\sqrt{1+\frac{\tau^{2}}{\kappa^{2}}C}\right)
+\left(\frac{\sqrt{1+\frac{\tau^{2}}{\kappa^{2}}C}-\frac{1}{\kappa}
P_{\tau}}{1+\frac{\tau^{2}}{\kappa^{2}}P^mP_m}\right) \otimes P_{\tau}
\nonumber
\\
\phantom{\Delta_{\tau} (P_{\tau} )=}
 -\frac{\tau^{2}}{\kappa}\left(\frac{\sqrt{1+\frac{\tau^{2}}{\kappa^{2}}C}-\frac{1}{\kappa}P_{\tau}}{1+\frac{\tau^2}{\kappa^{2}}
P^mP_m}\right)P^{j} \otimes P_{j},
\label{DPtau}
\\
\Delta_{\tau}(P_{i})=P_{i}\otimes \left(\frac{1}{\kappa}
P_{\tau}+\sqrt{1+\frac{\tau^{2}}{\kappa^{2}}C}\right) +1\otimes P_{i},
\\
\Delta_{\tau}(M_{ij})=M_{ij}\otimes 1+1\otimes M_{ij}, \qquad
i,j=1,\ldots,D-1,
\\
\Delta_{\tau}(M_{\tau i})=M_{\tau i}\otimes 1
 + \left(\frac{\sqrt{1+\frac{\tau^{2}}{\kappa^{2}}C}-\frac{1}{\kappa}
P_{\tau}}{1+\frac{\tau^{2}}{\kappa^{2}}P^mP_m}\right)\otimes M_{\tau i}
\nonumber
\\
\phantom{\Delta_{\tau}(M_{\tau i})=}
+\frac{\tau^{2}}{\kappa}\left(\frac{\sqrt{1+\frac{\tau^{2}}{\kappa^{2}}C}
-\frac{1}{\kappa}P_{\tau}}{1+\frac{\tau^{2}}{\kappa^{2}} P^mP_m}\right)P^{j} \otimes M_{ij},
\label{DMtau}
\end{gather}
where $\tau^{2}=g_{00}$ and after normalization it can be reduced to $\pm 1$ and
$C=P^{\mu}P_{\mu}=(\tau^2)^{-1}P^2_\tau+g^{i j}P_{i}P_{j}$,
$\mu,\nu=\tau,i$.
For this system of generators one has
\begin{gather}
\Pi_{\tau}=\frac{1}{\kappa}P_{\tau}+\sqrt{1+\frac{\tau^{2}}{\kappa^{2}}C},
\qquad
\Pi_{\tau}^{-1}=\frac{\sqrt{1+\frac{\tau^{2}}{\kappa^{2}}C}-\frac{1}{\kappa}P_{\tau}}{1+\frac{\tau^{2}}{\kappa^{2}}P^mP_m},
\label{Pi}
\\
C_{\tau}=\frac{\kappa^{2}}{\tau^{2}}\left(\Pi_{\tau}+\Pi_{\tau}^{-1}-2+\frac{\tau^{2}}{\kappa^{2}} P^mP_m
\Pi_{\tau}^{-1}\right)=\frac{2\kappa^{2}}{\tau^{2}}\left(\sqrt{1+\frac{\tau^{2}}{\kappa^{2}}C} -1\right),
\label{Casimir}
\end{gather}
as well as the following antipodes
\begin{gather}
S_{\tau}\left(P_{\tau}\right)
=-\left(P_{\tau}+\frac{\tau^{2}}{\kappa}\left(C+\frac{1}{2\kappa}P_{\tau}C_{\tau}\right)\right) \Pi_{\tau}^{-1},
\label{SPtau}
\\
S_{\tau}(P_{i})=-P_{i}\Pi_{\tau}^{-1},
\qquad
S_{\tau}(\Pi_{\tau})=\Pi_{\tau}^{-1},
\\
S_{\tau}(M_{ij})=-M_{ij},
\qquad
S_{\tau}(M_{\tau i})=-M_{\tau i}-\frac{\tau^{2}}{\kappa}P^{\alpha}M_{\alpha i}-\frac{\tau^{2}}{2\kappa^{2}}C_{\tau}M_{\tau i}.
\label{SMtau}
\end{gather}
The relations~\eqref{orthDec}--\eqref{SMtau} constitute the same Hopf algebra although written in another presentation
(system of generators).
The importance of such presentation has been shown in~\cite{BP_unif} by the relation to Majid--Ruegg
formulation~\cite{MR}.
Again, above coproducts and antipodes are formal power series in $\frac{1}{\kappa}$ as well.
It is to be observed that the elements $(P_\tau, \Pi_{\tau}, \Pi_{\tau}^{-1})$ are not algebraically independent, since
\begin{gather}
P_{\tau}=\frac{\kappa}{2}\left(\Pi_{\tau}-\Pi_{\tau}^{-1}\left(1+ \frac{\tau^{2}}{\kappa^{2}} P^mP_m\right)\right).
\label{d3}
\end{gather}
In the next section we show how to use this fact and by taking advantage of two (mutually inverse) group-like elements
$(\Pi_\tau, \Pi_\tau^{-1})$ one can eliminate inf\/inite power series from the formulas~\eqref{DPtau}--\eqref{SMtau}.

\textbf{$\boldsymbol{q}$-analog version of $\boldsymbol{U_{\kappa,\tau}(\iso(g))}$: $\boldsymbol{\tau^2\neq0}$.}
As a~f\/irst step, inside the Hopf algebra from the previous section\footnote{More generally, one can take, in fact, any
basis with $e_0=\tau$.
In such cases one has $\tau^2=g_{00}$ and $\tau_\mu=g_{0\mu}$.}, one can consider sub-Hopf algebra generated by elements
$(M_{ij}, P_{i}, M_{\tau i}, \Pi_{\tau}$, $\Pi_{\tau}^{-1})$ and call it $U_{q,\tau}(\iso(g))$.
Its generators satisfy the following relations\footnote{The sector $(M_{ij}, P_{i})$ remains standard,
cf.~formulas~\eqref{MM},~\eqref{PP}.}
\begin{gather}
\Pi_{\tau}\Pi_{\tau}^{-1}=1=\Pi_{\tau}\Pi_{\tau}^{-1},
\label{q1}
\\
 [P_{i},\Pi_{\tau} ]= [M_{ij},\Pi_{\tau} ]=0,
\qquad
 [M_{\tau i},\Pi_{\tau} ]=-\frac{i}{\kappa} P_{i},
\\
 [M_{\tau i,}P_{j} ]
=ig_{ij}\frac{\kappa}{2}\left(\Pi_{\tau}-\Pi_{\tau}^{-1}\left(1+\frac{\tau^{2}}{\kappa^{2}}P^m P_{m}\right)\right).
\label{q2}
\end{gather}
Commutators with $\Pi_{\tau}^{-1}$ can be easily calculated from the above
(e.g.\ $[M_{\tau i},\Pi_{\tau}^{-1}]=\frac{i}{\kappa}P_{i}\Pi_{\tau}^{-2}$).

Alternatively, one can abstractly def\/ine the algebraic structure of $U_{q,\tau}(\iso(g))$, in a~way similar
to the universal enveloping algebra, i.e.~as a~universal associative algebra generated by elements $(M_{ij},
P_{i}, M_{\tau i}, \Pi_{\tau}, \Pi_{\tau}^{-1})$ factorized by a~suitable (two-sided) ideal of
relations~\eqref{q1}--\eqref{q2}.
Coalgebraic structure
\begin{gather}
\Delta_{\tau}(\Pi_{\tau})=\Pi_{\tau}\otimes \Pi_{\tau},
\qquad
\Delta_{\tau}\big(\Pi_{\tau}^{-1}\big)=\Pi_{\tau}^{-1}\otimes \Pi_{\tau}^{-1},
\\
\Delta_{\tau}(P_{i})=P_{i}\otimes \Pi_{\tau}+1\otimes P_{i},
\qquad
i,j=1,\ldots,D-1,
\\
\Delta_{\tau}(M_{ij})=M_{ij}\otimes 1+1\otimes M_{ij},
\\
\Delta_{\tau}(M_{\tau i})=M_{\tau i}\otimes 1+\Pi_{\tau}^{-1}\otimes M_{\tau i}+\frac{\tau^{2}}{\kappa}P^{j}\Pi_{\tau}^{-1}\otimes M_{ij},
\label{q3}
\end{gather}
as well as antipodes
\begin{gather}
S_{\tau}(P_{i})=- P_{i}\Pi_{\tau}^{-1},
\qquad
S_{\tau}(\Pi^{\pm 1}_{\tau})=\Pi^{\mp 1}_{\tau},
\qquad
S_{\tau}(M_{ij})=-M_{ij},
\\
S_{\tau}(M_{\tau i})=-M_{\tau i}-\frac{\tau^{2}}{\kappa}
P^{k}M_{ki}-\frac{\tau^{2}}{2}\left(\Pi_{\tau}-\Pi_{\tau}^{-1}\left(1+ \frac{1}{\kappa^{2}}P^m P_{m}\right)\right)
M_{\tau i}-\frac{\tau^{2}}{2\kappa^{2}}C_{\tau}M_{\tau i}\!\!\!
\label{q4}
\end{gather}
can be calculated from~\eqref{DPtau}--\eqref{SMtau}.

In order to complete the def\/inition one leaves counit undeformed, i.e., $\epsilon (X)=0$ for
$X=(M_{ij},M_{\tau i},P_{i})$ and $\epsilon (\Pi_{\tau})=1=\epsilon (\Pi_{\tau}^{-1})$.
Deformed and undeformed central elements can now be expressed as
\begin{gather}
\label{q6}
C_{\tau}=\frac{\kappa^{2}}{\tau^{2}}\left(\Pi_{\tau}+\Pi_{\tau}^{-1}-2+\frac{\tau^{2}}{\kappa^{2}}P^m
P_{m}\Pi_{\tau}^{-1}\right),
\qquad
C=C_\tau\left(1+\frac{\tau^2}{4\kappa^2}C_\tau\right).
\end{gather}
It is important to note that the Hopf algebra $U_{q,\tau}(\iso(g))$ can be considered as an algebra over
polynomial ring $\mathbb{C}[{1\over\kappa}]$ (instead of $h$-adic ring $\mathbb{C}[[{1\over \kappa}]]$), since it
contains only polynomial expressions in the formal variable ${\frac{1}{\kappa}}$.
Because of this we are entitled to introduce a~new (non-isomorphic) Hopf algebra over $\mathbb{C}$ by assigning some
numerical (complex in general) value to the parameter~$\kappa$ in the formulas~\eqref{q1}--\eqref{q6}.
A~real value of~$\kappa$ is necessary if one wants to preserve the real form~\eqref{RC1} of the Hopf algebra.
This solves the so-called specialization problem for $U_{q,\tau}(\iso(g))$ in the case $\tau^2\neq0$.
First of all one can prove that dif\/ferent values of~$\kappa$ give rise to the same (isomorphic) Hopf algebras.
It can be seen if one changes the generators by the re-scaling $P_i\mapsto {\frac{1}{\kappa}}P_i$ which is equivalent
to setting $\kappa=1$\footnote{Some authors are used to use similar formulation in the bicrossproduct basis, see
e.g.~\cite{SitDur}.}.

One can argue that this new Hopf algebra has some advantages with respect to the previous ones.
For example, in the (semi-)simple case there is a~duality between Drinfel'd--Jimbo quantized enveloping algebras and some
matrix (or coordinate) Hopf algebras.
The same is expected for the case above~\cite{PW-2, PW-1}.

Moreover, one can also introduce an element $P_{\tau}$ expressed in terms of $q$-analog algebra ge\-ne\-ra\-tors by the
formula~\eqref{d3}.
This means that $U_{q,\tau}(\iso(g))$ contains as a~subalgebra the universal enveloping algebra $U(\iso(g))$.
However it is not a~sub-Hopf algebra.
This fact is meaningful for the representation theory of $U_{q,\tau}(\iso(g))$: any representation of
$U_{q,\tau}(\iso(g))$ becomes automatically a~representation of the corresponding Lie algebra $\iso(g)$.
The inverse statement is in general not true.
Instead one gets the following selection rule: a~representation of an orthogonal Lie algebra~$\iso(g)$ becomes at the
same time a~representation of its quantum version provided the elements $\Pi_\tau$, $\Pi_\tau^{-1}$ as given by the
formula~\eqref{Pi} are well def\/ined (self-adjoint) operators in the representation space and are mutually inverse of
each other.
From one hand, these might be complicated issues involving numerical value of the parameter~$\kappa$, domain of
$P_{\tau}$, etc.
From the other hand representations with a~constant value for the mass Casimir operator~$C$, e.g.\ irreducible ones, are very welcome.

From the point of view of applications in physical theories, specialization of the deformation parameter allows to
interpret it as some physical constant of Nature, e.g quantum gravity scale~$M_{\rm QG}$.
However the value of it depends on a~system of units one is using.
For example, one should be able to use natural (Planck) system of units, $\hbar=c=1$.
This f\/its very well with the re-scaling property mentioned above.

\subsection[The $D=2+(D-2)$ orthogonal decomposition: $\tau^2=0$]{The $\boldsymbol{D=2+(D-2)}$ orthogonal decomposition: $\boldsymbol{\tau^2=0}$}

\looseness=-1
In the previous subsection we were dealing with the $q$-analog version of the standard $\kappa$-defor\-ma\-tion
(i.e.~time-like case which includes the $\kappa$-Poincar\'{e} algebra).
For $\tau^2\neq0$ the corresponding \mbox{$r$-matrix} satisf\/ies MYBE.
It should be stressed that such $q$-analog version has not been considered for non-standard deformations (i.e.~for
which $r$-matrix satisf\/ies CYBE and the cocycle twist exists\footnote{It is known that for standard deformations of
semi-simple Lie algebra there exist a~cochain twist which determines a~weaker quasi-Hopf algebra structure on the
corresponding enveloping algebra instead.
In the case of non-semi-simple Poincar\'{e} Lie algebra the existence of such cochain twist has been proved
in~\cite{YZ}.
However our formulas for the coproduct \eqref{DPtau}--\eqref{DMtau} do not admit a~cochain twist for $\tau^2\neq0$ as
noticed in~\cite{BLP}.}).
In this section we will show that, in fact, it is possible to construct the $q$-analog for the case of $\tau^2=0$, which
includes the null-plane quantum Poincar\'{e} Hopf algebra~\cite{Ballesteros_null1,Ballesteros_null2,Ballesteros_null3}
(also known as the light-like deformation).
Throughout this section we assume non-Euclidean signature.

We remind the $D=2+(D-2)$ orthogonal decomposition for inhomogeneous orthogonal Hopf algebra of non-Euclidean signature
which relies on a~suitable choice of basis in the vector space~$V$.
This in turn invokes change of the generators of $U(\iso(g))$ which are now related with the so-called ``null-plane''
basis in $\iso(g)$
\begin{gather}
P_{\mu}= (P_{+},P_{-},P_{a} ),
\qquad
M_{\mu \nu}= (M_{+ -},M_{+ a},M_{- a},M_{ab} ),
\qquad
a,b=1,2,\ldots, D-2.
\label{lc1}
\end{gather}
Here we took two (non-orthogonal) null vectors $\tau^{\mu}\equiv \tau_{+}^{\mu}=(1,0,{\ldots}, 0)$,
$\tilde{\tau}^{\mu}\equiv \tau_{-}^{\mu}=(0,1,0,{\ldots}, 0)$: $\tau_{+}\tau_{-}=1$ (as spanning of 2-dimensional
Lorentzian subspace)\footnote{Starting from non-zero vector $\tau: \tau^2=0$, one can decompose the space
$V^{D}=V^{2}\times V^{D-2}$, by an appropriate choice of basis vectors, into the orthogonal product of 2-dimensional
Lorentzian space $\{V^{2},g_{AB}\}$ with a~$(D-2)$-dimensional one $\{V^{D-2},g_{ab}\}$: $(A,B=+,-)$,
$(a,b=1,2,\ldots, D-2)$.
Moreover, the total metric $g_{\mu \nu}=g_{AB}\times g_{ab}$ becomes a~product metric.
We choose $g_{AB}=\left(
\begin{matrix}
0 & 1
\\
1 & 0
\end{matrix}
\right)$ in its anti-diagonal (null-plane) form.} in order to obtain the convenient null-plane (a.k.a.\
light-cone) basis in the space of the Lie algebra generators~\eqref{lc1}.
This algebra consists of the following (non-vanishing) commutators
\begin{alignat}{3}
& [M_{+a},M_{-b}]=-i(M_{ab}+g_{ab}M_{+-}),
\qquad &&
[M_{\pm a},M_{\pm b}]=0,&
\label{M+aM-b}
\\
& [M_{\pm a},M_{bc}]=i(g_{ab}M_{\pm c}-g_{a c}M_{\pm b}),
\qquad &&
[M_{+-},M_{\pm a}]=\pm iM_{\pm a},
\label{MpmMbc}
\\
& [M_{+-},P_{\pm}]=\pm i P_{\pm},
\qquad &&
[M_{\pm a},P_{b}]=ig_{ab}P_{\pm}, &
\label{M_pmP}
\\
& [M_{\pm a},P_{\pm}]=[M_{+-},P_{a}]=0,
\qquad &&
[M_{\pm a},P_{\mp}]=-iP_{a} &
\label{M_pmP_pm}
\end{alignat}
together with the standard commutation relations within the $(D-2)$-dimensional
sector $(M_{ab}$,
$P_{a},g_{ab})$, cf.~\eqref{MM},~\eqref{PP}.

The universal formula for the coalgebra structure, in this case reduces to
\begin{gather}
\Delta_{\tau}\left(P\right)=P\otimes \left(1+\frac{1}{\kappa} P_{+}\right) +1\otimes P
\qquad
\text{for}
\quad
P\in \{P_{+},P_{a}\},
\label{DtauM1}
\\
\Delta_{\tau}(P_{-})=P_{-}\otimes \left(1+\frac{1}{\kappa} P_{+}\right)
+\left(1+\frac{1}{\kappa}P_{+}\right)^{-1}\otimes P_{-}
\nonumber
\\
\phantom{\Delta_{\tau}(P_{-})=}
 -\frac{1}{\kappa}\left(P_{-}+\frac{1}{2\kappa}C_{+}\right) \left(1+ \frac{1}{\kappa}P_{+}\right)^{-1}\otimes
P_{+}-\frac{1}{\kappa} P^{a}\left(1+\frac{1}{\kappa}P_{+}\right)^{-1}\otimes P_{a},
\\
\Delta_{\tau} (M_{+ -} )=M_{+ -}\otimes 1+\left(1+\frac{1}{\kappa}P_{+}\right)^{-1}\otimes
M_{+ -}-\frac{1}{\kappa}P^{a}\left(1+\frac{1}{\kappa}P_{+}\right)^{-1}\otimes M_{+ a},
\\
\Delta_{\tau} (M_{- a} )=M_{- a}\otimes 1+\left(1+\frac{1}{\kappa}P_{+}\right)^{-1}\otimes M_{- a}
\nonumber
\\
\phantom{\Delta_{\tau} (M_{- a} )=}
 -\frac{1}{\kappa}\left(P_{-}+\frac{1}{2\kappa}C_{+}\right) \left(1+ \frac{1}{\kappa}P_{+}\right)^{-1}\!\otimes
M_{+ a}-\frac{1}{\kappa} P^{b}\left(1+\frac{1}{\kappa}P_{+}\right)^{-1}\!\otimes M_{ba},\!\!\!
\\
\Delta_{\tau} (M )=M\otimes 1+1\otimes M
\qquad
\text{for}
\quad
M\in \{M_{+ a},M_{ab}\}
\label{DtauM2}
\end{gather}
and can be obtained by twisting of undeformed (primitive) coproducts by the extended Jordanian twist~\cite{BP_unif}
\begin{gather}
\mathcal{F}=\exp  (-iM_{+ -}\otimes\ln\Pi_{+} ) \exp\left(-\frac{i}{\kappa}M_{+ a}\otimes P^{a}\Pi_{+}^{-1}\right)
\nonumber
\\
\phantom{\mathcal{F}}
=\exp \left(-\frac{i}{\kappa}M_{+ a}\otimes P^{a}\right) \exp\left(-iM_{+ -}\otimes\ln\Pi_{+}\right),
\label{twist_LC}
\end{gather}
which satisf\/ies the two-cocycle condition, see also~\cite{KLM, Mudrov}.
Moreover, the twist is unitary, i.e.~the resulting deformation preserves the real form (cf.~\eqref{RC2}).
For this system of generators we have $\Pi_{+}\doteq 1+\frac{1}{\kappa}P_{+}$ and $C=2P_{+}P_{-}+P^{a}P_{a}$.
The antipodes are as follows
\begin{gather}
S_{\tau} (P )=-P\Pi_{+}^{-1}
\qquad
\text{for}
\quad
P\in \{P_{+},P_{a}\},
\\
S_{\tau}(P_{-})=-P_{-}\Pi_{+}-\frac{1}{\kappa}\left(1+ \frac{1}{2\kappa}P_{+}\right)P_{a}P^{a}\Pi_{+}^{-1},
\\
S_{\tau}(M_{+-})=-\Pi_{+}M_{+-}-\frac{1}{\kappa} P^{a}M_{+a},
\\
S_{\tau} (M_{-a} )=-\Pi_{+}M_{-a}-\frac{1}{\kappa}\left(P_{-}+\frac{1}{2\kappa}C\right)M_{+a}-\frac{1}{\kappa}P^{b}M_{ba},
\\
S_{\tau} (M )=-M
\qquad
\text{for}
\quad
M\in \{M_{+a},M_{ab}\}.
\end{gather}
Due to the presence of $\Pi^{-1}_+$ some of the above expressions are inf\/inite (formal) power series in~$\frac{1}{\kappa}$.
The Lie sub-algebra corresponding to the stability group of $\tau_{+}$ consists of $\iso(p-1,q-1)=\gen\{M_{ab},M_{+b}\}$,
i.e.~the generators with the primitive coproducts.
One can notice that~$P_+$ and~$\Pi_+$ are not algebraically independent and one can express one through
another: $P_{+}=\kappa(\Pi_+-1)$ which will be helpful in introducing the $q$-analog version of the above Hopf algebra.

\textbf{$\boldsymbol{q}$-analog version of $\boldsymbol{U_{\kappa,\tau}(\iso(g))}$: $\boldsymbol{\tau^2=0}$.}
Similarly to the previous section we start from considering the Hopf sub-algebra generated by the following
elements $(M_{+-},M_{+a},M_{-a},M_{ab},P_{a}$, $P_{-},\Pi_{+},\Pi_{+}^{-1})$ and we denote it as before as
$U_{q,\tau}(\iso(g))$.
The only dif\/ference is that now $\tau^2=0$ and one deals with non-standard (triangular) deformation.
It is generated by the following relations
\begin{gather}
 [M_{+ a},M_{- b} ]=-i (M_{ab}+g_{ab}M_{+ -} ),
\qquad
 [M_{\pm a},M_{\pm b} ]=0,
\label{Mpm_q}
\\
 [M_{\pm a},M_{bc} ]=i (g_{ab}M_{\pm c}-g_{ac}M_{\pm b} ),
\qquad
 [M_{+ -},M_{\pm a} ]=\pm iM_{\pm a},
\\
 [M_{+ -},\Pi_{+} ]=i (\Pi_{+}-1 ),
\qquad
 [M_{+ -},P_{-} ]=-iP_{-},
\qquad
 [M_{+ a},P_{b} ]=i\kappa g_{ab} (\Pi_{+}-1 ),
\\
 [M_{-a},P_{-} ]=0,
\qquad
 [M_{-a},P_{+} ]=-iP_{a},
\qquad
 [M_{-a},\Pi_{+} ]=-\frac{i}{\kappa}P_{a},
\\
 [M_{-a},P_{b} ]=ig_{ab}P_{-},
\qquad
 [M_{+a},\Pi_{+} ]= [M_{+a},P_{+} ]= [M_{+ -},P_{a} ]=0.
\end{gather}
The commutation relations within the $(D-2)$-dimensional sector $(M_{ab},P_{a},g_{ab})$ stay standard,
cf.~\eqref{MM},~\eqref{PP}.
Another way would be, to abstractly def\/ine the algebraic structure as universal unital and associative algebra over the
complex numbers generated by the elements $(M_{+-},M_{+a},M_{-a},M_{ab},P_{a},P_{-},\Pi_{+},\Pi_{+}^{-1})$
and factorized by the corresponding (two-sided) ideal of above relations.

Then a~coalgebra for the $q$-analog version in this case takes the form
\begin{gather}
\Delta_{\tau} (P_{a} )=P_{a}\otimes \Pi_{+}+1\otimes P_{a},
\\
\Delta_{\tau} (\Pi_{+} )=\Pi_{+}\otimes \Pi_{+},
\qquad
\Delta_{\tau} \big(\Pi_{+}^{-1} \big)=\Pi_{+}^{-1}\otimes \Pi_{+}^{-1},
\\
\Delta_{\tau}(P_{-})=P_{-}\otimes \Pi_{+}+\Pi_{+}^{-1}\otimes P_{-}-\left(P_{-}+\frac{1}{2\kappa}C\right)\Pi_{+}^{-1}
\otimes  (\Pi_{+}-1 ) -\frac{1}{\kappa}P^{a}\Pi_{+}^{-1}\otimes P_{a},\!\!\!
\\
\Delta_{\tau} (M_{+ -} )=M_{+ -}\otimes 1+\Pi_{+}^{-1}\otimes M_{+ -}-\frac{1}{\kappa}P^{a}\Pi_{+}^{-1}\otimes M_{+ a},
\\
\Delta_{\tau} (M_{- a} )=M_{- a}\!\otimes 1+\Pi_{+}^{-1}\!\otimes M_{- a}-\frac{1}{\kappa}\left(P_{-}\!+\frac{1}{2\kappa}C\right)\Pi_{+}^{-1}\!
\otimes M_{+ a}-\frac{1}{\kappa}P^{b}\Pi_{+}^{-1}\!\otimes M_{ba},\!\!\!
\\
\Delta_{\tau} (M )=M\otimes 1+1\otimes M
\qquad
\text{for}
\quad
M\in \{M_{+ a},M_{ab}\},
\end{gather}
where now $C=2\kappa (\Pi_{+}-1) P_{-}+P^{a}P_{a}$.
Including the following antipodes
\begin{gather}
S_{\tau}(P_{a})=-P_{a}\Pi_{+}^{-1},
\qquad
S_{\tau}(\Pi_{+})=\Pi_{+}^{-1},
\\
S_{\tau}(P_{-})=-P_{-}\Pi_{+}-\frac{1}{2\kappa}(1+\Pi_{+}) P_{a}P^{a}\Pi_{+}^{-1},
\\
S_{\tau}(M_{+-})=-\Pi_{+}M_{+-}-\frac{1}{\kappa} P^{a}M_{+a},
\\
S_{\tau}(M_{-a})=-\Pi_{+}M_{-a}-\frac{1}{\kappa}\left(P_{-}+\frac{1}{2\kappa}C\right)M_{+a}-\frac{1}{\kappa}P^{b}M_{ba},
\\
S_{\tau}(M)=-M
\qquad
\text{for}
\quad
M\in \{M_{+a},M_{ab}\}
\label{Spm_q}
\end{gather}
one completes the Hopf algebra structure provided the counits remain undeformed, i.e.~$\epsilon(X)=0$
for $X=(M_{+-},M_{+a},M_{-a},M_{ab},P_{a},P_{-},\Pi_{+},\Pi_{+}^{-1})$ and
$\epsilon(\Pi_+)=1=\epsilon(\Pi_+^{-1})$.

\looseness=-1
One can notice that analogously to the previous case the algebra $U_{q,\tau}(\iso(g))$ can be seen as an algebra over
polynomial ring $\mathbb{C}[\frac{1}{\kappa}]$.
By assigning the numerical value to~$\kappa$ in formulas~\eqref{Mpm_q}--\eqref{Spm_q} we are introducing new
(non-isomorphic) Hopf algebra over $\mathbb{C}$, which constitutes the so-called specialized ($q$-analog) version
$U_{q,\tau}(\iso(g))$ for the case of $\tau^2=0$.
Again, one can easily re-scale the momenta $P^{a}$, $P^{-}$ by $\frac{1}{\kappa}$, so in fact, one can prove that
dif\/ferent values of the parameter~$\kappa$ give rise to isomorphic Hopf-algebras.
Additionally, more general version of this construction is possible provided that one starts up from any basis in the
associate vector space with $e_0=\tau$.

From the considerations presented in this section one can deduce that any universal enve\-lo\-ping algebra twisted by an
(extended) Jordanian twist admits specialization of the deformation parameter to a~numerical value.

\subsection[$\kappa(\tau)$-Minkowski spacetime]{$\boldsymbol{\kappa(\tau)}$-Minkowski spacetime}

Quantum $\kappa$-Minkowski spacetime $\mathcal{M}_{\kappa,\tau}^{D}$ is usually def\/ined as an algebra (complex,
universal, unital and associative) generated by the following relations~\cite{koma95,LLM}
\begin{gather}
\left[{x}^{\mu},{x}^{\nu}\right]=\frac{i}{\kappa}\left(\tau^{\mu}{x}^{\nu}-\tau^{\nu} {x}^{\mu}\right)
\label{kMtau}
\end{gather}
of Lie algebra type\footnote{At the moment we consider ${\frac{1}{\kappa}}$ as a~formal variable.}, where $\tau^{\mu}$
are (real numerical) contravariant components of the vector $\tau \in V$ with respect to some basis $\{e_{\mu}\}_{\mu=0}^{D-1}$.

On one hand the algebra~\eqref{kMtau} is well adopted to the following action of the quantum
$U_{\kappa,\tau}(\iso(g))$
\begin{gather}
P_{\mu}\triangleright {x}^{\nu}=-\imath \delta_{\mu}^{\nu},
\qquad
M_{\mu \nu}\triangleright {x}^{\rho}=i\left(g_{\mu \alpha}\delta_{\nu}^{\rho}-g_{\nu \alpha}\delta_{\mu}^{\rho}\right){x}^{\alpha}.
\label{kM1}
\end{gather}
This implies $\Pi_{\tau}^{\pm 1}\triangleright {x}^{\mu}={x}^{\mu}\mp \frac{i}{\kappa}\tau^{\mu}$.
On the linear subspace spanned by the generators $\operatorname{lin}_\mathbb{C}\{x^1,\ldots$,
$x^D\}$ this action is
equivalent to the (complexif\/ied) vector representation~\eqref{rep}.
The way it extends to the polynomial expressions $x^{\mu_1}\cdots x^{\mu_k}$ is controlled by the coproduct.
Under the action the algebra~\eqref{kMtau} becomes a~covariant quantum space ($\equiv$ Hopf module algebra)
in a~sense of the compatibility condition (a.k.a.\
generalized Leibniz rule)
\begin{gather}
L\triangleright ({x}\cdot{y})=(L_{(1)}\triangleright {x})\cdot(L_{(2)}\triangleright {y})
\label{kM2}
\end{gather}
as well as
\begin{gather}
(L\cdot M) \triangleright {x}=L\triangleright \left(M\triangleright {x}\right)
\label{kM22}
\end{gather}
for any ${x}, {y}\in \mathcal{M}_{\kappa,\tau}^{D}$,
$L,M\in U_{\kappa,\tau}(\iso(g))$, where, for simplicity,
we have used Sweedler type notation for the coproduct: $\Delta_{\tau}(L)=L_{(1)}\otimes L_{(2)}$.
On the other hand, the presentation of the algebra~\eqref{kMtau} provides its natural real form by the requirement that
the generators $({x}^{\mu})^{\dagger}={x}^{\mu}$ are self-adjoint in analogy to~\eqref{RC1}.
The compatibility condition between two real structures can be expressed (see e.g.~\cite{Sitarz_star}) as a~reality
condition
\begin{gather}
L^{\ast}\triangleright {x}^{\dagger}=\big(S_{\tau}(L^{\ast})^{\ast}\triangleright {x}\big)^{\dagger}
\label{RC3}
\end{gather}
for the corresponding representation (module structure).
This property is enough to be checked on Hermitean generators: $L\in (M_{\mu \nu},P_{\rho})$, ${x}\in ({x}^{\mu})$ since
formula~\eqref{RC3} is consistent with both multiplications, i.e.~\eqref{kM2} and~\eqref{kM22}.

It should be observed that the metric components are not involved in the def\/inition of~\eqref{kMtau}, so the algebra is
independent of the metric itself and the metric signature in particular.
One can see it also by making use of a~general covariance in order to change the system of generators in (\ref{kMtau})
(cf.~\eqref{cov}).
Indeed, introducing new generators $\tilde x^\alpha=(A^{-1})^\alpha_\mu x^\mu$ and new components $\tilde
\tau^\alpha=(A^{-1})^\alpha_\mu \tau^\mu$ ($\tilde e_\alpha=A_\alpha^\mu e_\mu$, $A^\mu_\alpha\in {\rm GL}(D, \mathbb{R})$) one
preserves the form of~\eqref{kMtau} as well as the reality condition.
It shows that the real algebra~\eqref{kMtau} is, in fact, independent of the components of the vector $\tau\neq0$ (for
$\tau=0$ one obtains undeformed Abelian algebra).
In particular, one can always reach the well-known standard form of the $\kappa$-Minkowski spacetime algebra\footnote{To this aim we take any basis with $e_0=\tau$.}
\begin{gather}
\left[x^{0}, x^{i}\right]=\frac{i}{\kappa} x^{i},
\qquad
\left[x^{i}, x^{j}\right]=0,
\qquad
i,j=1,\ldots,D-1.
\label{kM3}
\end{gather}
One can conclude that up to the isomorphism mentioned above, for any dimension there is only one real $\kappa$-Minkowski
spacetime algebra $\mathcal{M}^D_{\kappa,\tau}$, which is covariant as a~Hopf module algebra with respect to the
action~\eqref{kM1} of dif\/ferent, in general, Hopf algebras equipped with dif\/ferent reality structures.
In fact, to be more precise, we have to distinguish three non-isomorphic options (incarnations): each one is adopted to
the corresponding form of quantum $U(\iso(g))$:

i) $\mathcal{M}^D_{\kappa,\tau}$ is closed in $h$-adic topology (see~\cite{BPsigma});

ii) $\mathcal{M}^D_{q,\tau}$ is considered as an algebra over polynomial ring $\mathbb{C}[{1\over\kappa}]$ admitting
only polynomial expressions in the formal variable ${\frac{1}{\kappa}}$ ($q$-analog version);

iii) we assign to~$\kappa$ numerical value.
In this case a~value of~$\kappa$ becomes irrelevant since it can be removed by re-scaling: $\tau^\mu\mapsto
{\frac{1}{\kappa}}\tau^\mu$ in~\eqref{kMtau} (or alternatively $x^0\mapsto \kappa x^0$ in~\eqref{kM3}).
Such algebra~$\mathcal{M}^D_{\tau}$ is isomorphic to the enveloping algebra of the solvable Lie algebra denoted usually\footnote{According to a~classif\/ication scheme~\cite{Graaf} for all 4-dimensional solvable Lie algebras (see
Appendix), we shall be using later, it is denoted as~$M^2$.} as~$\mathfrak{an}^{D}$. This version has found numerous
applications, e.g.\ from the point of view of the spectral triples~\cite{Andrea,Sitarz3,Sitarz2,Sitarz1} or in the group field
theories~\cite{Oriti} which are connected with loop quantum gravity and spin foams approach.

\subsection{Crossed product -- unif\/ied description for DSR algebras}

The property~\eqref{kM2} allows us to introduce a~larger algebra which unif\/ies $\iso(g)$ generators with that of
$\mathcal{M}^D_{\kappa,\tau}$ by making use of a~crossed product construction (see e.g.~\cite{BPsigma}).
This is the so-called DSR (deformed special relativity) algebra and it is based on the multiplication formula
\begin{gather*}
(f\otimes L)\rtimes (g\otimes M)=f(L_{(1)}\triangleright g)\otimes L_{(2)}M
\end{gather*}
providing the following crossed commutations
\begin{gather*}
 [1\otimes L,f\otimes 1 ]_{\rtimes}=(L_{(1)}\triangleright f)\otimes L_{(2)}-f\otimes L
\end{gather*}
between elements of two ingredient algebras: $\mathcal{M}_{\kappa,\tau}^{D}$ and
$\mathcal{U}_{\kappa,\tau}(\iso(g))$ with the action induced by~\eqref{kM1}.

Therefore the covariant form of DSR algebra, generalized now to any metric~$g$, any vector~$\tau$ and living in
arbitrary dimensions, obeys the algebraic relations~\eqref{MM},~\eqref{PP},
\eqref{kMtau} supplemented by the following
cross-commutation relations
\begin{gather*}
[{P}_{\mu},{x}^{\rho}]_{\rtimes}=-\imath \delta_{\mu}^{\rho}\Pi_{\tau}+i\frac{\tau_{\mu}}{\kappa}{P}^\rho,
\\
[{M}_{\mu \nu},{x}^{\rho}]_{\rtimes}=i(g_{\mu \alpha}\delta_{\nu}^{\rho}-g_{\nu
\alpha}\delta_{\mu}^{\rho}) {x}^{\alpha}+\frac{i}{\kappa}g^{\rho\alpha}(\tau_{\mu}{M}_{\alpha
\nu}-\tau_{\nu}{M}_{\alpha \mu}).
\end{gather*}
The obvious real form of this algebra (with Hermitean generators in the formulas above) is induced
from~\eqref{RC1},~\eqref{RC2} and~\eqref{RC3}.
It has been shown in~\cite{BPsigma} that deformed and undeformed (with commuting spacetime variables and primitive
coproducts, i.e.~for $\kappa\mapsto\infty$) DSR algebras are isomorphic each other (in general, such statement holds
true provided that one deals with twisted deformation~\cite{BPsigma}).

\section[Twist deformations of $\kappa$-Minkowski spacetime]{Twist deformations of $\boldsymbol{\kappa}$-Minkowski spacetime}\label{Section4}

As it is known, one can distinguish two types of quantum deformations of the corresponding universal enveloping
algebras: non-standard (triangular) provided by two-cocycle twist and standard, quasi-triangular one.
In the f\/irst case the corresponding classical $r$-matrix satisf\/ies Yang--Baxter equation with vanishing Schouten
brackets.
In the latter one deals with modif\/ied Yang--Baxter equation with invariant (non-trivial) Schouten brackets.
One of the advantages of the twist deformation is that it provides straightforwardly the universal~$R$-matrix and the
explicit formula for star-product, which is consistent with Hopf-algebraic actions.
Twisted deformations are especially useful, e.g.\ in gravity~\cite{Wess2,Aschieri1,Wess3,Aschieri3,Aschieri2},
f\/ield~\cite{ncqft1,ncqft3,ncqft4,ncqft2,ncqft6,ncqft7,ncqft8,ncqft5} and gauge~\cite{gauge3,gauge2, gauge1} theories on
noncommutative spaces and other applications requiring a~star-product formalism~\cite{star-pr}.
However it is known that $\kappa$-Minkowski spacetime with $\kappa$-Poincar\'{e} Hopf algebra as a~symmetry cannot be
obtained by twisting.
Nevertheless, both objects can be further quantized by twist.
Also some extensions of Poincar\'{e} algebra are amenable to twist formulation~\cite{BP1,Bu,star-pr}.
That is why the last part of this paper we devote to (further) twisting of $\kappa (\tau)$-deformed noncommutative
spacetimes.

We recall that twisting two-tensors~$F$ are invertible elements fulf\/illing 2-cocycle and norma\-li\-zation
conditions~\cite{ChP, Drinfeld1,Drinfeld3}.
Let us also remind that in the process of twisted deformation of the underlying spacetime algebra ($H$-module algebra)
the current `kappa' $*$-multiplication is replaced by a~new twist-deformed one
\begin{gather}
x\star_F y=m\circ F^{-1}\triangleright (x\otimes y)=(\bar{\mathrm{f}}^{\alpha}\triangleright x)\star
(\bar{\mathrm{f}}_{\alpha}\triangleright y),
\label{tsp}
\end{gather}
where $\triangleright$ denotes the classical action (e.g.\ like the one in (\ref{kM1})).
2-cocycle condition guarantees associativity of the corresponding twisted star-product~\eqref{tsp}.
Twisted deformations lead to noncommutative spacetimes which, in general form, involve dimensionfull parameters
\begin{gather*}
[{x}^{\mu}{x}^{\nu}]_{\star}=\Xi^{\mu \nu}({x})
=i\theta^{\mu \nu}+i\theta^{\mu \nu}_{\rho}{x}^{\rho}+i\theta^{\mu \nu}_{\lambda \rho}x^\lambda {x}^{\rho}+\cdots,
\end{gather*}
with a~constant, Lie-algebraic, quadratic, etc.
contributions.
Such deformation of spacetime algebra however, requires suitable modif\/ication of its relativistic symmetries as well.
According to the Leibniz rule~\eqref{kM2} the coalgebra sector (of the Hopf algebra) would change correspondingly.
In the case under consideration this can be done by twisting
\begin{gather}
\Delta_{\tau}^{F}(X)=F\Delta_{\tau}(X)F^{-1}.
\label{twisted_cop}
\end{gather}
In this section we will focus on Lie algebra type quantized noncommutative 4-dimensional spacetimes obtained from
twisting of $\kappa (\tau)$-Minkowski spacetime (\ref{kMtau})\footnote{Lie-algebraic deformations of undeformed
Minkowski spacetime algebra by Abelian twists from the Zakrzewski list~\cite{Zakrz-Comm} have been considered for the
f\/irst time in~\cite{LW}, see also~\cite{MD}.}.
After such twisting, it will become a~Hopf module algebra over the twisted $\kappa$-Poincar\'{e} Hopf algebra
$U_{\kappa,\tau}^{F}(\iso(g))$.
It is a~way to obtain new quantum algebras from the $\kappa (\tau)$-Minkowski one.

For this purpose we shall consider certain extensions of the classical $r$-matrices corresponding to $\kappa$-deformations.
Then using techniques proposed in~\cite{Tol0704.0081a,Tol0704.0081b} and~\cite{Lyakhovsky} one is able to write the
corresponding twists.
Zakrzewski~\cite{Zakrz-Comm} has already proposed a~list of Abelian extensions of $r_{\tau}$ which we shall use in the
time-, light- and space- like cases of the vector~$\tau$
\begin{gather*}
r_{\tau, {\rm ext}}=r_{\tau}+\xi P_{\tau}\wedge X,
\qquad
[P_{\tau},X]=0,
\end{gather*}
where~$X$ belongs to a~Lie algebra for the stability subgroup $G_{\tau}$ of $\tau$ (remembering that for time-like
case, $G_{\tau}={\rm SO}(3)$; for light-like case $G_{\tau}=E(2)={\rm ISO}(2)$; for space-like
case $G_{\tau}={\rm SO}(2,1)$~\cite{BP_unif}).
Here $\xi$ is a~new deformation parameter\footnote{Some further multi-parameter extensions are also possible
(see~\cite{Tol0704.0081a,Tol0704.0081b, Zakrz-Comm} for details).}.
Later on Lyakhovsky~\cite{Lyakhovsky} has found more sophisticated extensions of a~time-like $\kappa$-Poincar\'{e} case
(11~subcases)     showing at the same time that the list presented in~\cite{Zakrz-Comm} is incomplete (as already
mentioned by Zakrzewski himself).
According to our best knowledge the problem of f\/inal classif\/ication is still open.
We deform the multiplication in algebra~\eqref{kMtau} according to~\eqref{tsp} with the `classical' (undeformed)
action~\eqref{kM1}.
Our aim in this section is to describe new emerging spacetime algebras in terms of 4-dimensional Lie algebras.
Therefore one limits oneself to the case of Lie-algebraic deformations which for any value of the deformation parameter
turns out to be solvable.
Then we apply classif\/ication scheme for 4-dimensional
solvable Lie algebras as introduced in~\cite{Graaf}.
(For reader's convenience we recall basic facts of de~Graaf's approach in the Appendix.) In three cases we present also the
corresponding deformed $\kappa$-Poincar\'{e} coproducts.
Main results of this section are presented in the following Table~\ref{Table1}.

\begin{table}[h!]
\centering\footnotesize
\caption{}\label{Table1}
\vspace{1mm}

\begin{tabular}{|l|c|c|c|}
\hline
type & $r$-matrix & twist & algebra type
\\
\hline
\multicolumn{4}{|c|}{light-like case with $\tau^{+}=(1,0,0,0)$ and metric in $2+2$ decomposition\tsep{1pt}\bsep{1pt}}
\\
\hline
$\rm L1$ & $\xi P_{+}\wedge M_{+1}$ & $e^{i\xi\kappa (M_{+1}\wedge\ln\Pi_{+})}$ & $M_{a=1}^{3}$\tsep{3pt}
\\
$\rm L2$ & $\xi P_{+}\wedge M_{3}$
& $e^{i\xi\kappa (M_{3}\wedge\ln\Pi_{+})}$ & $M_{a,b}^{6}$ with $a=- \frac{(3+(2\kappa\xi )^{2})}{9}$,\\
&&& $b=\frac{(1+(2\kappa \xi )^{2})}{27}$\bsep{2pt}
\\
\hline
\multicolumn{4}{|c|}{space-like case with $\tau^{\mu}=(0,1,0,0)$ and $\eta_{\mu \nu}=(+,-,-,-)$\tsep{1pt}\bsep{1pt}}
\\
\hline
$\rm S1$ & $\xi P_{1}\wedge M_{1}$ & $e^{i\xi P_{1}\otimes M_{1}}$ & $M_{a,b}^{6} \ \text{with} \  a=-\frac{(3+(\kappa
\xi )^{2})}{9}$,\tsep{3pt}\\
&&& $b=\frac{(1+(\kappa \xi )^{2})}{27}$
\\
$\rm S2$ & $\xi P_{1}\wedge (M_{1}+N_{3})$ & $e^{i\xi P_{1}\otimes (M_{1}+N_{3})}$ & the same as
${\rm S1}$
\\
$\rm S3$ & $\xi P_{1}\wedge N_{3}$ & $e^{i\xi P_{1}\otimes N_{3}}$ & the same as ${\rm S1}$\bsep{1pt}
\\
\hline
\multicolumn{4}{|c|}{time-like case with $\tau=(1,0,0,0)$ and $\eta_{\mu \nu}=(-,+,+,+)$\tsep{1pt}\bsep{1pt}}
\\
\hline
$\rm T1$ & $\xi M_{3}\wedge P_{0}$ & $e^{i\xi\kappa\ln\Pi_{0}\wedge M_{3}}$ & $M^6_{a,b}$ with
$a=-\frac{(3+4(\alpha \kappa)^{2})}{9}$,\tsep{3pt}\\
&&& $b=\frac{(1+4(\alpha \kappa)^{2})}{27}$
\\
$\rm T3$ & $\pm \frac{1}{2\kappa}M_{3}\wedge P_{0}+\xi\tilde M_{\pm}\wedge \tilde P_{\pm}$ & $e^{\xi\tilde
P_{\pm}\Pi_{0}^{\frac{1}{2}}\otimes\tilde M_{\pm}}e^{\pm \frac{i}{2}\ln \Pi_{0}\otimes M_{3}}$ &
$M^{13}_{b=-\frac{2}{9}}$ (as complex algebra)
\\
$\rm T4$ & $\pm \frac{1}{\kappa}M_{3}\wedge P_{0}$ &
$e^{\pm \xi\tilde M_{\pm}e^{-\sigma_\pm -\ln \Pi_{0}}\otimes P_{3}}e^{\sigma_\pm \otimes M_{3}}e^{\pm i\ln
\Pi_{0}\otimes M_{3}}$ & $M_8$ (as complex algebra)
\\
& $\pm \xi (P_{3}\wedge \tilde M_{\pm}+ M_{3}\wedge \tilde P_{\pm})$\tsep{1pt} & &\\
\hline
\end{tabular}
\end{table}

The last two cases come from Lyakhovsky f\/indings.
In his notation $\tilde P_\pm=P_1\pm iP_2$,
$\tilde M_\pm=M_1\pm i M_2$ and $\sigma_\pm=\ln(1+ \xi\tilde P_\pm)$.
This implies that these twists are not unitary and the corresponding spacetime algebras are complex.
Another fact is that the case $\rm S3$ with a~special values $\xi=\pm \frac{1}{\kappa}$, $\pm\frac{2}{\kappa}$ provide
nonequivalent deformations~\cite{Zakrz-Comm}.
The same is true for $\rm T1$ with $\xi=\pm {1\over 2\kappa}$, $\pm\frac{1}{\kappa}$, $\pm\frac{2}{\kappa}$~\cite{Lyakhovsky}.

\looseness=-1
{\bf Twisting light-cone deformation.}
For convenience we choose a~light-cone basis (see footnote~9).
Two Abelian twists corresponding to the cases denoted as $\rm L1$ and $\rm L2$ have the form\footnote{Notice that both
commuting elements $M_{+1}$, $\ln\Pi_{+}$ have primitive coproducts in the light-cone deformed $\kappa$-Poincar\'{e}
algebra.
The same is true for $M_{3}$, $\ln\Pi_{+}$.}
\begin{gather*}
F_{\rm L1}=e^{i\xi\kappa  (M_{+1}\wedge\ln\Pi_{+} )},
\qquad
F_{\rm L2}=e^{i\xi\kappa  (M_{3}\wedge\ln\Pi_{+} )}.
\end{gather*}
For these two examples, we shall demonstrate how by a~chain of consecutive linear transformations of generators one can
reach a~canonical form from~\cite{Graaf}.

L1.~From twist one obtains the $\star $-commutators as def\/ining relations for the algebra under consideration\footnote{We shall always write only non-vanishing commutators.}
\begin{gather*}
\left[x^{+},x^{1}\right]_{\star}=\frac{i}{\kappa}x^{1}+2i\xi x^{-},
\qquad
\left[x^{+},x^{2}\right]_{\star}=\frac{i}{\kappa}x^{2},
\qquad
\left[x^{+},x^{-}\right]_{\star}=\frac{i}{\kappa}x^{-}.
\end{gather*}
One can check that for any f\/ixed (real) value of the parameters $\kappa$, $\xi$ this is a~solvable Lie algebra.
Firstly we notice that the coordinates $(x^{1},x^{2},x^{-})$ make a~$L^{1}$ Abelian 3-dimensional
subalgebra.
Thus 4-dimensional algebra can be classif\/ied as $M_{a=1}^{3}$ in the following way:
\begin{itemize}\itemsep=0pt
\item[1.] Firstly we rescale $x^{+}$ as $\frac{\kappa}{i}x^{+}=\tilde{x}^{0}$ to obtain
$[\tilde{x}^{0},x^{1}]=x^{1}+2\kappa \xi x^{-}$,
$[\tilde{x}^{0},x^{2}]=x^{2}$,
$[\tilde{x}^{0},x^{-}]=x^{-}$.
\item[2.] Then change the generators as $\tilde{x}^{1}=x^{1}+\beta x^{-}$ and $\tilde{x}^{-}=x^{1}+\gamma x^{-}$ to get
$[\tilde{x}^{0},\tilde{x}^{1}]=\tilde{x}^{-}$ with $\gamma=(2\kappa \xi +\beta )$ and $[\tilde{x}^{0},x^{2}]=x^{2}$,
together with $[\tilde{x}^{0},\tilde{x}^{-}]=-\tilde{x}^{1}+2\tilde{x}^{-}$.
\end{itemize}

And this algebra can be classif\/ied as (cf.~Appendix)
\begin{gather*}
M_{a=1}^{3}: \ \left[x^{0},x^{1}\right]=x^{3},
\qquad
\left[x^{0},x^{2}\right]=x^{2},
\qquad
\left[x^{0},x^{3}\right]=-ax^{1}+ (1+a ) x^{3}
\qquad
\text{for}
\quad
a=1.
\end{gather*}
One can observe that the f\/inal form does not depend on the numerical values of the deformation parameters in this case.

L2.~The algebra obtained from the second twist is def\/ined by the following relations
\begin{gather*}
\left[x^{+},x^{1}\right]_{\star}=\frac{i}{\kappa}x^{1}-2i\xi x^{2},
\qquad
\left[x^{+},x^{2}\right]_{\star}=\frac{i}{\kappa}x^{2}+2i\xi x^{1},
\qquad
\left[x^{+},x^{-}\right]_{\star}=\frac{i}{\kappa}x^{-}.
\end{gather*}
Firstly we recognize its 3-dimensional subalgebra $(x^{-},x^{1},x^{2})$ as $L^{1}$-Abelian Lie algebra.
The whole 4-dimensional one undergoes the following changes:
\begin{itemize}\itemsep=0pt
\item[1.] $x^{+}$ goes into $\frac{\kappa}{i}x^{+}=x^{0}$, we also denote $x^{-}$ by $x^{3}$ and put $\alpha=2\kappa\xi$.
This way we get
\begin{gather*}
\left[x^{0},x^{1}\right]_{\star}=x^{1}-\alpha x^{2},
\qquad
\left[x^{0},x^{2}\right]_{\star}=x^{2}+\alpha x^{1},
\qquad
\left[x^{0},x^{3}\right]_{\star}=x^{3},
\\
\left[x^{3},x^{1}\right]_{\star}=0,
\qquad
\left[x^{3},x^{2}\right]_{\star}=0,
\qquad
\left[x^{1},x^{2}\right]_{\star}=0.
\end{gather*}
\item[2.] We rename the generators as $x^{1}\rightarrow x^{2}+\alpha x^{1}=\tilde{x}^{1}$, $\alpha\neq0$ and we get
\begin{gather*}
\left[x^{0},\tilde{x}^{1}\right]_{\star}=2\tilde{x}^{1}-\big(1+\alpha^{2}\big) x^{2},
\qquad
\left[x^{0},x^{2}\right]_{\star}=\tilde{x}^{1},
\qquad
\left[x^{0},x^{3}\right]_{\star}=x^{3}.
\end{gather*}
\item[3.] Subsequently we take  $\tilde{x}^{1}\rightarrow \tilde{x}^{1}+x^{3}=\bar{x}^{1}$,
$x^{2}\rightarrow x^{2}+x^{3}=\bar{x}^{2}$,
\begin{gather*}
\left[x^{0},\bar{x}^{1}\right]_{\star}=2\bar{x}^{1}+\alpha^{2}x^{3}-\left(1+\alpha^{2}\right) \bar{x}^{2},
\qquad
\left[x^{0},\bar{x}^{2}\right]_{\star}=\bar{x}^{1},
\qquad
\left[x^{0},x^{3}\right]_{\star}=x^{3}.
\end{gather*}
\item[4.] Once more introducing new generators: $x^{3}\rightarrow 2\bar{x}^{1}+\alpha^{2}x^{3}-(1+\alpha^{2})
\bar{x}^{2}=\bar{x}^{3}$ one obtains
\begin{gather*}
\left[x^{0},\bar{x}^{1}\right]_{\star}=\bar{x}^{3},
\qquad
\left[x^{0}, \bar{x}^{2}\right]_{\star}=\bar{x}^{1},
\qquad
\left[x^{0},\bar{x}^{3}\right]_{\star}=3\bar{x}^{3}+\left(1+\alpha^{2}\right) \bar{x}^{2}-\left(3+\alpha^{2}\right)\bar{x}^{1}.
\end{gather*}
\item[5.]
After f\/inal change  $x^{0}\rightarrow \frac{x^{0}}{3}=x^{0}$,
$\bar{x}^{1}\rightarrow 3\bar{x}^{1}=x^{1}$,
$\bar{x}^{2}\rightarrow 9\bar{x}^{2}=x^{2}$, $\bar{x}^{3}=x^{3}$ which leads~to
\begin{gather*}
\left[x^{0},x^{1}\right]_{\star}=x^{3},
\qquad
\left[x^{0},x^{2}\right]_{\star}=x^{1},
\qquad
\left[x^{0},x^{3}\right]_{\star}=x^{3}+\frac{\left(1+\alpha^{2}\right)}{27}x^{2}-\frac{\left(3+\alpha^{2}\right)}{9}x^{1}
\end{gather*}
\end{itemize}
with $\alpha=2\kappa \xi$.
The resulting algebra can be recognized as
\begin{gather*}
M_{a,b}^{6}:\left[x^{0},x^{1}\right]=x^{3},
\qquad
\left[x^{0},x^{2}\right]=x^{1},
\qquad
\left[x^{0},x^{3}\right]=ax^{1}+bx^{2}+x^{3}
\end{gather*}
with $a=-\frac{(3+(2\kappa \xi )^{2})}{9}$, $b=\frac{(1+(2\kappa \xi
)^{2})}{27}$.
This time isomorphism class of the resulting algebra does depend on the numerical value of the product $\kappa\xi$.

{\bf Twisting of the symmetries.}
As we mentioned in the previous part one should also perform a~twisting of the coalgebra sector of $\kappa$-Poincar\'{e}
Hopf algebra via~\eqref{twisted_cop}.
In this section we will focus on 4-dimensional quantized Poincar\'{e} Hopf algebra $U_{\kappa,\tau}(\iso(1,3))$.
Firstly we will use the light-like twists $F_{\rm L1}$ and $F_{\rm L2}$ and subsequently one of the time-like twist $F_{\rm T1}$.

{\bf Twisting of null-plane Poincar\'{e}.}
Let us focus now on the quantum null-plane Poincar\'{e} algebra case~\cite{Ballesteros_null1,Ballesteros_null2,Ballesteros_null3}.
As reminded in the previous sections such algebra is a~result of twist deformation.
The classical $r$-matrix corresponding to the vector $\tau_{+}$ reads
\begin{gather*}
r_{\rm LC}=M_{+ -}\wedge P_{+}+M_{+ a}\wedge P^{a}
\end{gather*}
and $\tau_{+}^{2}=0$ so it satisf\/ies CYBE~\eqref{z2}.
The coproducts $\Delta_{\rm LC}(X)$ obtained directly from the twist~\eqref{twist_LC} via (\ref{twisted_cop})
$\Delta_{\rm LC}(X)=\mathcal{F}\Delta_{0}(X)\mathcal{F}^{-1}$ are related with the universal
ones~\eqref{DtauM1}--\eqref{DtauM2} via $\mathcal{R}\Delta_{\rm LC}(X)\mathcal{R}^{-1}=\Delta_{\rm LC}^{\rm op}(X)=\Delta_{\tau}(X)$, where
$\mathcal{R}=\mathcal{F}_{21}\mathcal{F}^{-1}$ is a~triangular quantum~$R$-matrix.

The algebra relations are~\eqref{M+aM-b}--\eqref{M_pmP_pm}.

{\bf Twist $\boldsymbol{F_{\rm L1}}$.}
The twisted deformation~\eqref{twisted_cop} of the coproducts $\Delta_{\rm LC}(X)$ with the twist $F_{\rm L1}$ (see
the table) results in the following coalgebra
\begin{gather*}
\Delta_{\rm L1}(P_{+})=P_{+}\otimes 1+\Pi_{+}\otimes P_{+},
\\
\Delta_{\rm L1} (P_{a} )
=P_{a}\otimes 1+\Pi_{+}\otimes P_{a}-\xi (g_{1a}P_{+}\otimes \kappa\ln\Pi_{+}-\Pi_{+}\kappa\ln\Pi_{+}\otimes g_{1a}P_{+} ),
\\
\Delta_{\rm L1}(P_{-})=P_{-}\otimes \Pi_{+}^{-1}+\Pi_{+}\otimes P_{-}-\frac{1}{\kappa}P_{+}\otimes
\left(P_{-}+\frac{1}{2\kappa}C_{+}\right) \Pi_{+}^{-1}-\frac{1}{\kappa}P_{a}\otimes P^{a}\Pi_{+}^{-1}
\\
\phantom{\Delta_{\rm L1}(P_{-})=}{}
 +\xi\big(P_{1}\otimes \Pi_{+}^{-1} (\kappa\ln\Pi_{+} ) -\Pi_{+} (\kappa\ln\Pi_{+} ) \otimes P_{1}\big)
\\
\phantom{\Delta_{\rm L1}(P_{-})=}{}
 -\frac{\xi^{2}}{2}\big(P_{+}\otimes\Pi_{+}^{-1} (\kappa\ln\Pi_{+} )^{2}+ (\kappa\ln\Pi_{+} )^{2}\Pi_{+}\otimes P_{+}\big)
\\
\phantom{\Delta_{\rm L1}(P_{-})=}{}
 +\frac{\xi}{\kappa}\big(P_{+}\otimes (\kappa\ln\Pi_{+} ) P_{1}\Pi_{+}^{-1}-P_{1} (\kappa\ln\Pi_{+} )
 \otimes P_{+}\Pi_{+}^{-1}\big)
\\
\phantom{\Delta_{\rm L1}(P_{-})=}{}
 +\frac{\xi}{\kappa}P_{+} (\kappa\ln\Pi_{+} ) \otimes
P_{1}\Pi_{+}^{-1}+\frac{\xi^{2}}{2\kappa}P_{+} (\kappa\ln\Pi_{+} )^{2}\otimes P_{+}\Pi_{+}^{-1},
\\
\Delta_{\rm L1} (M_{- 1} )=M_{- 1}\otimes \Pi_{+}^{-1}+1\otimes M_{- 1}-\frac{1}{\kappa}M_{21}\otimes
P_{2}\Pi_{+}^{-1}+M_{+-}\otimes \Pi_{+}^{-1} (\xi\kappa\ln\Pi_{+} )
\\
\phantom{\Delta_{\rm L1} (M_{- 1} )=}{}
 - (\xi\kappa\ln\Pi_{+} )\otimes M_{+-}-\xi M_{+1}\otimes P_{1}\Pi_{+}^{-1}+\xi P_{1}\Pi_{+}^{-1}\otimes \Pi_{+}^{-1}M_{+1}
\\
\phantom{\Delta_{\rm L1} (M_{- 1} )=}{}
  +M_{+1}\otimes \Pi_{+}^{-1}\left(-\frac{1}{\kappa}\left(P_{-}+\frac{1}{2\kappa}C_{+}\right) +\frac{1}{2}\left(\xi\kappa\ln\Pi_{+}\right)^{2}\right)
\\
\phantom{\Delta_{\rm L1} (M_{- 1} )=}{}
  +\xi\ln\Pi_{+}M_{+ 1}\otimes P_{1}\Pi_{+}^{-1}+\frac{1}{2} (\xi\kappa\ln\Pi )^{2}\frac{1}{\kappa}M_{+1}\otimes P_{+}\Pi_{+}^{-1}
\\
\phantom{\Delta_{\rm L1}(M_{-1})=}
  +\frac{1}{2} (\xi\kappa\ln\Pi_{+} )^{2}\otimes M_{+1}-\frac{1}{\kappa}M_{+2}\otimes  (\xi\kappa\ln\Pi_{+} ) P_{2}\Pi_{+}^{-1},
\\
\Delta_{\rm L1} (M_{- 2} )=M_{-2}\otimes \Pi_{+}^{-1}+1\otimes M_{-2}-\frac{1}{\kappa}M_{12}\otimes
P_{1}\Pi_{+}^{-1}-\xi (\kappa\ln\Pi_{+} ) \otimes M_{12}
\\
\phantom{\Delta_{\rm L1} (M_{- 2} )=}{}+\xi M_{12}\otimes \Pi_{+}^{-1} (\kappa\ln\Pi_{+} )
 -\xi M_{+1}\otimes P_{2}\Pi_{+}^{-1}+\xi P_{2}\Pi_{+}^{-1}\otimes \Pi_{+}^{-1}M_{+1}
\\
\phantom{\Delta_{\rm L1} (M_{- 2} )=}{}
 -M_{+2}\otimes \Pi_{+}^{-1}\left(\frac{1}{2} (\xi\kappa\ln\Pi_{+} )^{2}+\frac{1}{\kappa}\left(P_{-}+\frac{1}{2\kappa} C_{+}\right)\right)
\\
\phantom{\Delta_{\rm L1} (M_{- 2} )=}{}
 +\frac{1}{2\kappa} (\xi\kappa\ln\Pi )^{2}M_{+2}\otimes P_{+}\Pi_{+}^{-1}-\frac{1}{2} (\xi\kappa \ln
\Pi_{+} )^{2}\otimes M_{+2}
\\
\phantom{\Delta_{\rm L1} (M_{- 2} )=}{}
 +\xi M_{+2}\otimes\ln\Pi_{+}P_{1}\Pi_{+}^{-1}-\xi\ln\Pi_{+}M_{12}\otimes P_{+}\Pi_{+}^{-1}+\xi\ln\Pi_{+}M_{+2}\otimes P_{1}\Pi_{+}^{-1},
\\
\Delta_{\rm L1} (M_{+b} )=M_{+b}\otimes 1+1\otimes M_{+b},
\qquad
b=1,2,
\\
\Delta_{\rm L1}(M_{3})=M_{3}\otimes 1+1\otimes M_{3}-\xi (M_{+2}\otimes \kappa\ln\Pi_{+}-\kappa\ln\Pi_{+}\otimes M_{+2} ),
\\
\Delta_{\rm L1} (M_{+ -} )=M_{+ -}\otimes \Pi_{+}^{-1}+1\otimes M_{+ -}-\frac{1}{\kappa}M_{+ a}\otimes P^{a}\Pi_{+}^{-1}+M_{+1}\otimes \Pi_{+}^{-1}\xi\kappa\ln\Pi_{+}
\\
\phantom{\Delta_{\rm L1} (M_{+ -} )=}{}
-\xi\kappa\ln\Pi_{+}\otimes M_{+1}
 -\xi P_{+}\Pi_{+}^{-1}\otimes M_{+1}\Pi_{+}^{-1}-\frac{1}{\kappa} M_{+ 1} (\xi\kappa\ln\Pi_{+}-1 )\otimes P_{+}\Pi_{+}^{-1}.
\end{gather*}
This constitutes the twisted algebra $U_{\kappa,\tau}^{\rm L1}(\iso(1,3))$.

One should notice that after twisting we are still able to construct $q$-analog version.
It is due to the fact that above expressions are polynomial in the new parameter $\xi$ as well.
It agrees with the previous observation that the corresponding spacetime algebra type does not depend on $\xi$.

Another remark is that $\lim\limits_{\kappa\rightarrow \infty}(\kappa\ln\Pi_{+})=P_{+}$.
This allows us to calculate easily the limit $\Delta_{0,L1}=\lim\limits_{\kappa\rightarrow \infty}\Delta_{\rm L1}$.

\textbf{Twist $\boldsymbol{F_{\rm L2}}$.}
The corresponding unitary twist to the classical $r$-matrix $r_{\rm L2}$ is\footnote{Abelian form of the twist is due to
similar reasons as explained previously.}
\begin{gather*}
F_{\rm L2}=\exp  (i\xi\kappa M_{3}\wedge\ln\Pi_{+} ).
\end{gather*}

The (further) twisting via relation~\eqref{twisted_cop} of the deformed coalgebra structure $\Delta_{\rm LC}$ results in
\begin{gather*}
\Delta_{\rm L2}(P_{1})=\left(\Pi_{0}-1+\frac{1}{2}\big(\Pi_{+}^{i\xi\kappa}+\Pi_{+}^{-i\xi\kappa}\big)
\right) \otimes P_{1}+P_{1}\otimes \frac{1}{2}\big(\Pi_{+}^{i\xi\kappa}+\Pi_{+}^{-i\xi\kappa}\big)
\\
\phantom{\Delta_{\rm L2}(P_{1})}{}
 -\frac{1}{2}\big(\Pi_{+}^{i\xi\kappa}-\Pi_{+}^{-i\xi\kappa}\big) \otimes iP_{2}+iP_{2}\otimes
\frac{1}{2}\big(\Pi_{+}^{i\xi\kappa}-\Pi_{+}^{-i\xi\kappa}\big),
\\
\Delta_{\rm L2}(P_{2})= (\Pi_{0}-1 ) \otimes P_{2}+P_{2}\otimes \frac{1}{2}\big(\Pi_{+}^{i\xi\kappa}+\Pi_{+}^{-i\xi\kappa}\big)
+\frac{1}{2}\big(\Pi_{+}^{i\xi\kappa}+\Pi_{+}^{-i\xi\kappa}\big) \otimes P_{2}
\\
\phantom{\Delta_{\rm L2}(P_{2})=}{}
 -iP_{1}\otimes \frac{1}{2}\big(\Pi_{+}^{i\xi\kappa}-\Pi_{+}^{-i\xi\kappa}\big)
+\frac{1}{2}\big(\Pi_{+}^{i\xi\kappa}-\Pi_{+}^{-i\xi\kappa}\big) \otimes iP_{1},
\\
\Delta_{\rm L2}(P_{+})=P_{+}\otimes 1+\Pi_{+}\otimes P_{+},
\\
\Delta_{\rm L2}(P_{-})=P_{-}\otimes \Pi_{+}^{-1}+\Pi_{+}\otimes P_{-}-\frac{1}{\kappa}P_{+}\otimes
\left(P_{-}+\frac{1}{2\kappa}C\right) \Pi_{+}^{-1}
\\
\phantom{\Delta_{\rm L2}(P_{-})=}{}
 -\frac{1}{\kappa}P_{a}\otimes P^{a}\Pi_{+}^{-1}\frac{1}{2}\big(\Pi_{+}^{i\xi\kappa}+\Pi_{+}^{-i\xi\kappa}\big)
+\frac{i}{\kappa}P_{1}\frac{1}{2}\big(\Pi_{+}^{i\xi\kappa}-\Pi_{+}^{-i\xi\kappa}\big) \otimes P_{2}\Pi_{+}^{-1}
\\
\phantom{\Delta_{\rm L2}(P_{-})=}{}
-\frac{i}{\kappa}P_{2} \frac{1}{2}\big(\Pi_{+}^{i\xi\kappa}-\Pi_{+}^{-i\xi\kappa}\big) \otimes P^{1}\Pi_{+}^{-1}
-\frac{1}{\kappa}P_{1}\left(-1+\frac{1}{2}\big(\Pi_{+}^{i\xi\kappa}+\Pi_{+}^{-i\xi\kappa}\big)\right) \otimes P_{2}\Pi_{+}^{-1}
\\
\phantom{\Delta_{\rm L2}(P_{-})=}{}
 -\frac{1}{\kappa}P_{2}\left(-1+\frac{1}{2}\big(\Pi_{+}^{i\xi\kappa}+\Pi_{+}^{-i\xi\kappa}\big)\right) \otimes P_{2}\Pi_{+}^{-1}
 +\frac{i}{\kappa}P_{1}\otimes P_{2}\Pi_{+}^{-1}\frac{1}{2}\big(\Pi_{+}^{i\xi\kappa}-\Pi_{+}^{-i\xi\kappa}\big)
\\
\phantom{\Delta_{\rm L2}(P_{-})=}{}
-\frac{i}{\kappa} P_{2}\otimes P_{1}\Pi_{+}^{-1}\frac{1}{2}\big(\Pi_{+}^{i\xi\kappa}-\Pi_{+}^{-i\xi\kappa}\big),
\\
\Delta_{\rm L2}(M_{ab})=\Delta_{0}(M_{ab}),
\qquad
a,b=1,2,
\qquad
\alpha,\beta=1,2,
\\
\Delta_{\rm L2}(M_{+-})=M_{+ -}\otimes \Pi_{+}^{-1}+1\otimes M_{+ -}
-\frac{1}{\kappa}M_{+ \alpha}\otimes \frac{1}{2}\big(\Pi_{+}^{i\xi\kappa}+\Pi_{+}^{-i\xi\kappa}\big)
P^{\alpha}\Pi_{+}^{-1}
\\
\phantom{\Delta_{\rm L2}(M_{+-})=}{}
 -\frac{1}{\kappa}\left(-1+\frac{1}{2}\big(\Pi_{+}^{i\xi\kappa}+\Pi_{+}^{-i\xi\kappa}\big)\right)
M_{+ \alpha}\otimes P^{\alpha}\Pi_{+}^{-1}
\\
\phantom{\Delta_{\rm L2}(M_{+-})=}{}
 +\frac{i}{2\kappa}\big(\Pi_{+}^{i\xi\kappa}-\Pi_{+}^{-i\xi\kappa}\big) M_{+ 1}\otimes
P_{2}\Pi_{+}^{-1}-\frac{i}{2\kappa} M_{+1}\otimes \big(\Pi_{+}^{i\xi\kappa}-\Pi_{+}^{-i\xi\kappa}\big)P_{2}\Pi_{+}^{-1}
\\
\phantom{\Delta_{\rm L2}(M_{+-})=}{}
  +\frac{i}{2\kappa}M_{+ 2}\otimes \big(\Pi_{+}^{i\xi\kappa}-\Pi_{+}^{-i\xi\kappa}\big)
P_{1}\Pi_{+}^{-1}-\frac{i}{2\kappa}\big(\Pi_{+}^{i\xi\kappa}-\Pi_{+}^{-i\xi\kappa}\big) M_{+2}\otimes P_{1}\Pi_{+}^{-1},
\\
\Delta_{\rm L2}(M_{+1})=M_{+ 1}\otimes \frac{1}{2}\big(\Pi_{+}^{i\xi\kappa}+\Pi_{+}^{-i\xi\kappa}\big)
+\frac{1}{2}\big(\Pi_{+}^{i\xi\kappa}+\Pi_{+}^{-i\xi\kappa}\big) \otimes M_{+1}
\\
\phantom{\Delta_{\rm L2}(M_{+1})=}{}
 +\frac{1}{2}\big(\Pi_{+}^{i\xi\kappa}-\Pi_{+}^{-i\xi\kappa}\big) \otimes iM_{+2}-iM_{+2}\otimes
\frac{1}{2}\big(\Pi_{+}^{i\xi\kappa}-\Pi_{+}^{-i\xi\kappa}\big),
\\
\Delta_{\rm L2}(M_{+2})=M_{+ 2}\otimes \frac{1}{2}\big(\Pi_{+}^{i\xi\kappa}+\Pi_{+}^{-i\xi\kappa}\big)
+\frac{1}{2}\big(\Pi_{+}^{i\xi\kappa}+\Pi_{+}^{-i\xi\kappa}\big) \otimes M_{+2}
\\
\phantom{\Delta_{\rm L2}(M_{+2})=}{}
 +\frac{1}{2}\big(\Pi_{+}^{i\xi\kappa}-\Pi_{+}^{-i\xi\kappa}\big) \otimes iM_{+1}-iM_{+1}\otimes
\frac{1}{2}\big(\Pi_{+}^{i\xi\kappa}-\Pi_{+}^{-i\xi\kappa}\big),
\\
\Delta_{\rm L2}(M_{-1})=\frac{1}{2}\big(\Pi_{+}^{i\xi\kappa}+\Pi_{+}^{-i\xi\kappa}\big) \otimes
M_{-1}+M_{- 1}\otimes \Pi_{+}^{-1}\frac{1}{2}\big(\Pi_{+}^{i\xi\kappa}+\Pi_{+}^{-i\xi\kappa}\big)
\\
\phantom{\Delta_{\rm L2}(M_{-1})=}{}
 +\frac{1}{2}\big(\Pi_{+}^{i\xi\kappa}-\Pi_{+}^{-i\xi\kappa}\big) \otimes iM_{-2}-iM_{-2}\otimes
\frac{1}{2}\big(\Pi_{+}^{i\xi\kappa}-\Pi_{+}^{-i\xi\kappa}\big) \Pi_{+}^{-1}
\\
\phantom{\Delta_{\rm L2}(M_{-1})=}{}
 +i\frac{1}{\kappa}M_{+2}\otimes \frac{1}{2}\big(\Pi_{+}^{i\xi\kappa}-\Pi_{+}^{-i\xi\kappa}\big)
\left(P_{-}+\frac{1}{2\kappa}C\right) \Pi_{+}^{-1}
\\
\phantom{\Delta_{\rm L2}(M_{-1})=}{}
 -\frac{1}{\kappa}M_{+1}\otimes \frac{1}{2}\big(\Pi_{+}^{i\xi\kappa}+\Pi_{+}^{-i\xi\kappa}\big)
\left(P_{-}+\frac{1}{2\kappa}C\right) \Pi_{+}^{-1}
\\
\phantom{\Delta_{\rm L2}(M_{-1})=}{}
 +\frac{1}{\kappa}M_{21}\frac{1}{2}\big(\Pi_{+}^{i\xi\kappa}-\Pi_{+}^{-i\xi\kappa}\big) \otimes
iP_{1}\Pi_{+}^{-1}-\frac{1}{\kappa} M_{21}\otimes \frac{1}{2}\big(\Pi_{+}^{i\xi\kappa}+\Pi_{+}^{-i\xi\kappa}\big)P_{2}\Pi_{+}^{-1}
\\
\phantom{\Delta_{\rm L2}(M_{-1})=}{}
 +\xi\big(P_{+}\Pi_{+}^{-1}\otimes M_{3}\Pi_{+}^{-1}-M_{3}\otimes P_{+}\Pi_{+}^{-1}\big),
\\
\Delta_{\rm L2}(M_{-2})=M_{-2}\otimes \frac{1}{2}\big(\Pi_{+}^{i\xi\kappa}+\Pi_{+}^{-i\xi\kappa}\big)
\Pi_{+}^{-1}+\frac{1}{2} \big(\Pi_{+}^{i\xi\kappa}+\Pi_{+}^{-i\xi\kappa}\big) \otimes M_{-2}
\\
\phantom{\Delta_{\rm L2}(M_{-2})=}{}
 +iM_{-1}\otimes \frac{1}{2}\big(\Pi_{+}^{i\xi\kappa}-\Pi_{+}^{-i\xi\kappa}\big)
\Pi_{+}^{-1}-\frac{1}{2}\big(\Pi_{+}^{i\xi\kappa}-\Pi_{+}^{-i\xi\kappa}\big) \otimes iM_{-1}
\\
\phantom{\Delta_{\rm L2}(M_{-2})=}{}
 -\frac{1}{\kappa}M_{+2}\otimes \frac{1}{2}\big(\Pi_{+}^{i\xi\kappa}+\Pi_{+}^{-i\xi\kappa}\big)
\left(P_{-}+\frac{1}{2\kappa}C\right) \Pi_{+}^{-1}
\\
\phantom{\Delta_{\rm L2}(M_{-2})=}{}
 -\frac{i}{\kappa}M_{+ 1}\otimes \frac{1}{2}\big(\Pi_{+}^{i\xi\kappa}-\Pi_{+}^{-i\xi\kappa}\big)
\left(P_{-}+\frac{1}{2\kappa}C\right) \Pi_{+}^{-1}
\\
\phantom{\Delta_{\rm L2}(M_{-2})=}{}
 -\frac{1}{\kappa}M_{3}\otimes \frac{1}{2}\big(\Pi_{+}^{i\xi\kappa}+\Pi_{+}^{-i\xi\kappa}\big)
P_{1}\Pi_{+}^{-1}-\frac{i}{\kappa}M_{3} \frac{1}{2}\big(\Pi_{+}^{i\xi\kappa}-\Pi_{+}^{-i\xi\kappa}\big) \otimes P_{2}\Pi_{+}^{-1}
\\
\phantom{\Delta_{\rm L2}(M_{-2})=}{}
 -\xi\left(M_{3}\otimes P_{2}\Pi_{+}^{-1}-P_{2}\Pi_{+}^{-1}\otimes M_{3}\Pi_{+}^{-1}\right).
\end{gather*}
Here we notice that after twisting we are unable to construct $q$-analog version.
It is caused by the fact that above expressions are formal power series in the new parameter $\xi$ as well.
The isomorphism class of the corresponding spacetime
algebra does depend on the value of $\xi\kappa$.

Since $\lim\limits_{\kappa\rightarrow\infty} \Pi_0^{\pm\kappa}=\exp (\pm P_0)$ we are able to calculate the limit
$\Delta_{0,{\rm L2}}=\lim\limits_{\kappa\rightarrow\infty}\Delta_{\rm L2}$ by replacing
\begin{gather*}
\Pi_0^{\pm 1}\mapsto 1,
\qquad
{\frac{1}{2}}\big(\Pi_{0}^{i\xi\kappa}+\Pi_{0}^{-i\xi\kappa}\big)\mapsto \cos(\xi P_0),
\qquad
{\frac{1}{2}}\big(\Pi_{0}^{i\xi\kappa}-\Pi_{0}^{-i\xi\kappa}\big)\mapsto i\sin(\xi P_0)
\end{gather*}
and dropping out all terms proportional to ${\frac{1}{\kappa}}$.

{\bf Twisting of~$\boldsymbol{\kappa}$-Poincar\'{e}.}
The algebra part of the twisted $U_{\kappa,\tau}^{\rm T1}(\iso(1,3))$ will be still described by~\eqref{MM},~\eqref{PP},
but
the coalgebra part will be deformed accordingly to~\eqref{twisted_cop}.
By de\-for\-ming~\eqref{DPtau}--\eqref{DMtau} with $F_{\rm T1}$ we obtain\footnote{Here the following standard notation for
Lorentz rotations is used: $M_{i}=\frac{1}{2} \epsilon_{ijk}M_{jk}$.}
\begin{gather*}
\Delta_{\rm T1}(P_{0})=P_{0}\otimes \Pi_{0}+\Pi_{0}^{-1}\otimes
P_{0}-\frac{\tau^{2}}{\kappa}P_{3}\Pi_{0}^{-1}\otimes P_{3} -i\frac{\tau^{2}}{\kappa}P_{1}\Pi_{0}^{-1}\frac{1}{2}\big(\Pi_{0}^{i\xi\kappa}-\Pi_{0}^{-i\xi\kappa}\big) \otimes
P_{2}
\\
\phantom{\Delta_{\rm T1}(P_{0})=}{}
+i\frac{\tau^{2}}{\kappa}P_{2}\Pi_{0}^{-1}\frac{1}{2}\big(\Pi_{0}^{i\xi\kappa}-\Pi_{0}^{-i\xi\kappa}\big)\otimes P_{1}
+i\frac{\tau^{2}}{\kappa}P_{2}\Pi_{0}^{-1}\otimes \frac{1}{2}\big(\Pi_{0}^{i\xi\kappa}-\Pi_{0}^{-i\xi\kappa}\big)
P_{1}\\
\phantom{\Delta_{\rm T1}(P_{0})=}{}
-iP_{1}\frac{\tau^{2}}{\kappa}\Pi_{0}^{-1}\otimes \frac{1}{2}\big(\Pi_{0}^{i\xi\kappa}-\Pi_{0}^{-i\xi\kappa}\big) P_{2}
 -\frac{\tau^{2}}{\kappa}P^{\alpha}\Pi_{0}^{-1}\frac{1}{2}\big(\Pi_{0}^{i\xi\kappa}+\Pi_{0}^{-i\xi\kappa}\big)\otimes P_{\alpha}
\\
\phantom{\Delta_{\rm T1}(P_{0})=}{}
-\frac{\tau^{2}}{\kappa}P^{\alpha}\Pi_{0}^{-1}\otimes \left[\frac{1}{2} \big(\Pi_{0}^{i\xi\kappa}+\Pi_{0}^{-i\xi\kappa}\big)-1\right] P_{\alpha},
\qquad
\alpha,\beta=1,2,
\\
\Delta_{\rm T1}(P_{1})=P_{1}\otimes \left(\Pi_{0}-1\right) +P_{1}\otimes \frac{1}{2}\big(\Pi_{0}^{i\xi\kappa}+\Pi_{0}^{-i\xi\kappa}\big)
+\frac{1}{2}\big(\Pi_{0}^{i\xi\kappa}+\Pi_{0}^{-i\xi\kappa}\big) \otimes P_{1}
\\
\phantom{\Delta_{\rm T1}(P_{1})=}{}
 +\frac{i}{2}\big(\Pi_{0}^{i\xi\kappa}-\Pi_{0}^{-i\xi\kappa}\big) \otimes P_{2}-P_{2}
\otimes\frac{i}{2}\big(\Pi_{0}^{i\xi\kappa}-\Pi_{0}^{-i\xi\kappa}\big),
\\
\Delta_{\rm T1}(P_{2})=P_{2}\otimes \left(\Pi_{0}-1\right) +P_{2}\otimes \frac{1}{2}\big(\Pi_{0}^{i\xi\kappa}+\Pi_{0}^{-i\xi\kappa}\big)
+\frac{1}{2}\big(\Pi_{0}^{i\xi\kappa}+\Pi_{0}^{-i\xi\kappa}\big) \otimes P_{2}
\\
\phantom{\Delta_{\rm T1}(P_{2})=}{}
 +iP_{1}\otimes \frac{1}{2}\big(\Pi_{0}^{i\xi\kappa}-\Pi_{0}^{-i\xi\kappa}\big)
-i\frac{1}{2}\big(\Pi_{0}^{i\xi\kappa}-\Pi_{0}^{-i\xi\kappa}\big)\otimes P_{1},
\\
\Delta_{\rm T1}\left(P_{3}\right)=P_{3}\otimes \Pi_{0}+1\otimes P_{3},
\\
\Delta_{\rm T1}(M_{1})=M_{1}\otimes \frac{1}{2}\big(\Pi_{0}^{i\xi\kappa}+\Pi_{0}^{-i\xi\kappa}\big)
+\frac{1}{2}\big(\Pi_{0}^{i\xi\kappa}+\Pi_{0}^{-i\xi\kappa}\big) \otimes M_{1}
\\
\phantom{\Delta_{\rm T1}(M_{1})=}{}
 +\frac{i}{2}\big(\Pi_{0}^{i\xi\kappa}-\Pi_{0}^{-i\xi\kappa}\big) \otimes M_{2}-M_{2}\otimes
\frac{i}{2}\big(\Pi_{0}^{i\xi\kappa}-\Pi_{0}^{-i\xi\kappa}\big),
\\
\Delta_{\rm T1}(M_{2})=M_{2}\otimes \frac{1}{2}\big(\Pi_{0}^{i\xi\kappa}+\Pi_{0}^{-i\xi\kappa}\big)
+\frac{1}{2}\big(\Pi_{0}^{i\xi\kappa}+\Pi_{0}^{-i\xi\kappa}\big) \otimes M_{2}
\\
\phantom{\Delta_{\rm T1}(M_{2})=}{}
  +\frac{i}{2}\big(\Pi_{0}^{i\xi\kappa}-\Pi_{0}^{-i\xi\kappa}\big) \otimes M_{1}-M_{1}\otimes
\frac{i}{2}\big(\Pi_{0}^{i\xi\kappa}-\Pi_{0}^{-i\xi\kappa}\big),
\\
\Delta_{\rm T1}(M_{3})=M_{3}\otimes 1+1\otimes M_{3},
\\
\Delta_{\rm T1}(M_{01})=M_{01}\otimes \frac{1}{2}\big(\Pi_{0}^{i\xi\kappa}+\Pi_{0}^{-i\xi\kappa}\big)
+\Pi_{0}^{-1}\frac{1}{2} \big(\Pi_{0}^{i\xi\kappa}+\Pi_{0}^{-i\xi\kappa}\big) \otimes M_{01}
\\
\phantom{\Delta_{\rm T1}(M_{01})=}{}
 +M_{3}\Pi_{0}^{-1}\otimes \xi\tau^{2}P_{1}\Pi_{0}^{-1}-\tau^{2}\xi P_{1}\Pi_{0}^{-1}\otimes M_{3}
\\
\phantom{\Delta_{\rm T1}(M_{01})=}{}
 -M_{02}\otimes \frac{i}{2}\big(\Pi_{0}^{i\xi\kappa}-\Pi_{0}^{-i\xi\kappa}\big)
+\Pi_{0}^{-1}\frac{i}{2}\big(\Pi_{0}^{i\xi\kappa}-\Pi_{0}^{-i\xi\kappa}\big) \otimes M_{02}
\\
\phantom{\Delta_{\rm T1}(M_{01})=}{}
 +\frac{\tau^{2}}{\kappa}P_{2}\Pi_{0}^{-1}\otimes \frac{1}{2}\big(\Pi_{0}^{i\xi\kappa}+\Pi_{0}^{-i\xi\kappa}\big)
M_{3}-\frac{\tau^{2}}{\kappa}P_{3}\Pi_{\tau}^{-1}\frac{1}{2}\big(\Pi_{0}^{i\xi\kappa}+\Pi_{0}^{-i\xi\kappa}\big)\otimes M_{2}
\\
\phantom{\Delta_{\rm T1}(M_{01})=}{}
 +i\frac{\tau^{2}}{\kappa}P_{1}\Pi_{0}^{-1}\otimes \frac{1}{2}\big(\Pi_{0}^{i\xi\kappa}-\Pi_{0}^{-i\xi\kappa}\big)
M_{3}+i\frac{\tau^{2}}{\kappa}\frac{1}{2}\big(\Pi_{0}^{i\xi\kappa}-\Pi_{0}^{-i\xi\kappa}\big)P_{3}\Pi_{\tau}^{-1}\otimes M_{1},
\\
\Delta_{\rm T1}(M_{02})=M_{02}\otimes \frac{1}{2}\big(\Pi_{0}^{i\xi\kappa}+\Pi_{0}^{-i\xi\kappa}\big)
+\Pi_{0}^{-1}\frac{1}{2} \big(\Pi_{0}^{i\xi\kappa}+\Pi_{0}^{-i\xi\kappa}\big) \otimes M_{02}
\\
\phantom{\Delta_{\rm T1}(M_{02})=}{}
  -\tau^{2}\xi P_{2}\Pi_{0}^{-1}\otimes M_{3}+\tau^{2}\xi M_{3}\Pi_{0}^{-1}\otimes P_{2}\Pi_{0}^{-1}
\\
\phantom{\Delta_{\rm T1}(M_{02})=}{}
  +M_{01}\otimes \frac{i}{2}\big(\Pi_{0}^{i\xi\kappa}-\Pi_{0}^{-i\xi\kappa}\big)-\frac{i}{2}\big(\Pi_{0}^{i\xi\kappa}-\Pi_{0}^{-i\xi\kappa}\big)
\Pi_{0}^{-1}\otimes M_{01}
\\
\phantom{\Delta_{\rm T1}(M_{02})=}{}
 +\frac{\tau^{2}}{\kappa}P_{3}\Pi_{\tau}^{-1}\otimes \frac{1}{2}\big(\Pi_{0}^{i\xi\kappa}+\Pi_{0}^{-i\xi\kappa}\big)M_{1}
+i\frac{\tau^{2}}{\kappa}\frac{1}{2}\big(\Pi_{0}^{i\xi\kappa}-\Pi_{0}^{-i\xi\kappa}\big)P_{3}\Pi_{\tau}^{-1}\otimes M_{2}
\\
\phantom{\Delta_{\rm T1}(M_{02})=}{}
 +i\frac{\tau^{2}}{\kappa}P_{2}\Pi_{\tau}^{-1}\otimes M_{3}\frac{1}{2} \big(\Pi_{0}^{i\xi\kappa}-\Pi_{0}^{-i\xi\kappa}\big)
 -\frac{\tau^{2}}{\kappa}P_{1}\Pi_{\tau}^{-1}\otimes M_{3}\frac{1}{2}\big(\Pi_{0}^{i\xi\kappa}+\Pi_{0}^{-i\xi\kappa}\big),
\\
\Delta_{\rm T1}(M_{03})=M_{03}\otimes 1+\Pi_{0}^{-1}\otimes M_{03}-\xi\tau^{2} P_{3}\Pi_{0}^{-1}\otimes
M_{3}+ \xi\tau^{2} M_{3}\Pi_{0}^{-1}\otimes P_{3}\Pi_{0}^{-1}
\\
\phantom{\Delta_{\rm T1}(M_{03})=}{}
 -\frac{\tau^{2}}{\kappa}P_{2}\Pi_{0}^{-1}\otimes \frac{1}{2}\big(\Pi_{0}^{i\xi\kappa}+\Pi_{0}^{-i\xi\kappa}\big)
M_{1}-i\frac{\tau^{2}}{\kappa}P_{1}\Pi_{0}^{-1}\otimes \frac{1}{2}\big(\Pi_{0}^{i\xi\kappa}-\Pi_{0}^{-i\xi\kappa}\big) M_{1}
\\
\phantom{\Delta_{\rm T1}(M_{03})=}{}
 -i\frac{\tau^{2}}{\kappa}\frac{1}{2}\big(\Pi_{0}^{i\xi\kappa}\!-\Pi_{0}^{-i\xi\kappa}\big) P_{1}\Pi_{0}^{-1}\!\otimes
M_{1}-\frac{\tau^{2}}{\kappa}\left[\frac{1}{2}\big(\Pi_{0}^{i\xi\kappa}\!+\Pi_{0}^{-i\xi\kappa}\big) -1\right]\!
P_{2}\Pi_{0}^{-1}\!\otimes M_{1}
\\
\phantom{\Delta_{\rm T1}(M_{03})=}{}
 +\frac{\tau^{2}}{\kappa}\frac{1}{2}\big(\Pi_{0}^{i\xi\kappa}+\Pi_{0}^{-i\xi\kappa}\big) P_{1}\Pi_{0}^{-1}\otimes
M_{2}-i\frac{\tau^{2}}{\kappa}\frac{1}{2}\big(\Pi_{0}^{i\xi\kappa}-\Pi_{0}^{-i\xi\kappa}\big)P_{2}\Pi_{0}^{-1}\otimes M_{2}
\\
\phantom{\Delta_{\rm T1}(M_{03})=}{}
 -i\frac{\tau^{2}}{\kappa}P_{2}\Pi_{0}^{-1}\!\otimes \frac{1}{2}\big(\Pi_{0}^{i\xi\kappa}\!-\Pi_{0}^{-i\xi\kappa}\big)
M_{2}+\frac{\tau^{2}}{\kappa}P_{1}\Pi_{0}^{-1}\!\otimes \left[\frac{1}{2}\big(\Pi_{0}^{i\xi\kappa}\!+\Pi_{0}^{-i\xi\kappa}\big) -1\right]\! M_{2}.
\end{gather*}
One can compare the above results with the ones in the bicrossproduct basis included in~\cite{LL-0406155}.

Again by the similar to the previous rules one can calculate $\lim\limits_{\kappa\rightarrow \infty} \Delta_{\rm T1}$.

Now the special values $\xi=\pm {\frac{1}{2\kappa}}$, $\pm\frac{1}{\kappa}$, $\pm\frac{2}{\kappa}$ provide six new
deformations for which, in contrast to the general case, specialization of~$\kappa$ is possible.

\pdfbookmark[1]{Appendix}{app}
\section*{Appendix~A.~Some classes of 4-dimensional solvable Lie algebras}

For readers convenience we partially summarize the recent classif\/ication results of~\cite{Graaf} on which our
description is based (see also~\cite{Patera1,Patera2,Patera3} for earlier results and broader context).
For the sake of completeness we begin by recalling the def\/inition.
For given Lie algebra $\mathfrak{g}$ we def\/ine a~sequence of subalgebras of $\mathfrak{g}$ (the so-called lower derived
series) by $\mathfrak{g}_{(0)}=\mathfrak{g}$, $\mathfrak{g}_{(1)}=[\mathfrak{g}_{(0)},\mathfrak{g}_{(0)}],\dots$,
$\mathfrak{g}_{(i)}=[\mathfrak{g}_{(i-l)},\mathfrak{g}_{(i-l)}]$.
We call $\mathfrak{g}$ solvable if $\mathfrak{g}_{(n)}=0$ for some f\/inite~$n$.
In a~similar manner, the upper sequence $\mathfrak{g}^{(0)}=\mathfrak{g}$,
$\mathfrak{g}^{(1)}=[\mathfrak{g},\mathfrak{g}^{(0)}],\dots$,
$\mathfrak{g}^{(i)}=[\mathfrak{g},\mathfrak{g}^{(i-l)}]$
determines nilpotent Lie algebras.
For example, nilpotent (e.g.\
Abelian) algebras are solvable, whereas semisimple algebras are def\/initely nonsolvable.
Moreover, a~f\/inite-dimensional
Lie algebra $\mathfrak{g}$ over a~f\/ield of characteristic zero is solvable if and only if
$\mathfrak{g}_{(1)}\equiv \mathfrak{g}^{(1)}$ is nilpotent.

The strategy for the classif\/ication of all 4-dimensional
solvable Lie algebras undertaken in~\cite{Graaf} is based on the following observation.
Any $n$-dimensional
solvable Lie algebra~$L$
over a~f\/ield $\mathbb{F}$ admits a~presentation as $L=\mathbb{F}D\oplus K$,
where~$K$ is a~solvable Lie algebra of dimension $n-1$ and~$D$ is a~derivation of~$K$.
Moreover~$D$ is an outer derivation for non-Abelian~$L$.

Therefore one should f\/irstly know corresponding candidates for such subalgebras.
Classif\/ication of all 3-dimensional
real Lie algebras is well known for a~long time since Bianchi (Lie himself had
earlier classif\/ied the complex ones).
Here we repeat after~\cite{Graaf} all non-isomorphic classes of solvable ones\footnote{We write down only nonzero Lie brackets.}:
\begin{enumerate}\itemsep=0pt
\item[] $L^{1}$: the Abelian Lie algebra;

\item[] $L^{2}$: $[x^{3},x^{1}]=x^{1}$, $[x^{3},x^{2}]=x^{2}$ (3-dimensional $\kappa$-Minkowski spacetime algebra);

\item[] $L_{a}^{3}$: $[x^{3},x^{1}]=x^{2}$, $[x^{3},x^{2}]=ax^{1}+x^{2}$, where $a\in\mathbb{R}$ (or $\mathbb{C}$);

\item[] $L_{a}^{4}$: $[x^{3},x^{1}]=x^{2}$, $[x^{3},x^{2}]=ax^{1}$, where $a=0, 1,-1$ in the real case (or $a=0,1$ in the complex one).
\end{enumerate}

One should notice that only two of them $L^1$ and $L^4_0$ (Heisenberg Lie algebra) are nilpotent.

Thus 4-dimensional solvable Lie algebras are classif\/ied by adding derivations to the algebras listed above.
The f\/inal classif\/ication is done for arbitrary f\/ield $\mathbb{F}$ and can be found in~\cite{Graaf}.
Here we are interested only in the real (or complex) cases.
We are listing only those equivalence classes which are important in the context of extended $\kappa$-Minkowski
spacetime algebras studied in this paper\footnote{We do not know if the remaining classes can be obtained by twisting.}.
Following~\cite{Graaf} isomorphism classes are denoted by $M^{i}_a$, $i=1,\ldots, 14$ with a~suitable (discrete or
continuous) subscript~$a$ (below $x^0$ is a~derivation of $K=\gen\{x^1, x^2, x^3\}$):
\begin{enumerate}\itemsep=0pt
\item[I.] $K=L^{1}$:
\begin{enumerate}\itemsep=0pt
\item[i)] $M^{2}$, where $[x^{0},x^{1}]=x^{1}$, $[x^{0},x^{2}]=x^{2}$, $[x^{0},x^{3}]=x^{3}$
(4-dimensional $\kappa$-Minkowski spacetime algebra);

\item[ii)] $M_{a}^{3}$: $[x^{0},x^{1}]=x^{1}$, $[x^{0},x^{2}]=x^{3}$, $[x^{0},x^{3}]=-ax^{2}+(a+1) x^{3}$, where $a\in\mathbb{R}$ (or $\mathbb{C}$);

\item[iii)] $M_{a,b}^{6}$: $[x^{0},x^{1}]=x^{3}$, $[x^{0},x^{2}]=x^{1}$,
$[x^{0},x^{3}]=x^{3}+ax^{2}+bx^{1}$, where $a, b\in\mathbb{R}$ (or $\mathbb{C}$).
\end{enumerate}
\item[II.]
$K=L^{2}$:
\begin{enumerate}\itemsep=0pt
\item[i)] $M^{8}$: $[x^{1},x^{2}]=x^{2}$, $[x^{0},x^{3}]=x^{3}$.
\end{enumerate}
\item[III.] $K=L_{0}^{4}$:
\begin{enumerate}\itemsep=0pt
\item[i)] $M_{b}^{13}$: $[x^{0},x^{1}]=x^{1}+ b x^3$, $[x^{0},x^{2}]=x^{2}=[x^{3},x^{1}]$, $[x^{0},x^{3}]=x^{1}$,
where $b\in\mathbb{R}$ (or $\mathbb{C}$).
\end{enumerate}
\end{enumerate}

One can notice that algebra $M^8$ is a~direct sum of two 2-dimensional $\kappa$-Minkowski algebras.
It turns out (by Gr\"{o}bner basis analysis) that as a~complex Lie algebra it is isomorphic to the family of Lie
algebras denoted as $K_v$ in~\cite{Graaf}, where
$K_v$: $[x^{0},x^{1}]=x^{1}+v x^{2}$, $[x^{0},x^{2}]=x^{1}=[x^{3},x^{1}]$, $[x^{3},x^{2}]=x^{2}$, $v\in\mathbb{C}$.

\subsection*{Acknowledgements}

\looseness=-1
We are grateful to V.~Lyakhovsky for collaboration and discussions during the early stages of the work presented in
Section~\ref{Section4}.
We are also indebted to J.~Lukierski for critical remarks and pointing out the reference~\cite{LW}.
We would like to thank to the anonymous referees for relevant suggestions to improve the paper.
This work is a~part of the Polish National Science Centre (NCN) project 2011/01/B/ST2/03354.
AB acknowledges the f\/inancial support from FSS Mobility and Training Program as well as the hospitality of the Science
Institute of University of Iceland.

\pdfbookmark[1]{References}{ref}
\LastPageEnding

\end{document}